\documentclass[twocolumn,showpacs,floatfix]{revtex4}

\usepackage{graphics}

\begin{document}

\title{Squeezed-light generation in a nonlinear planar waveguide with a periodic corrugation}

\author{Jan Pe\v{r}ina, Jr.}
\author{Ond\v{r}ej Haderka}
\affiliation{Joint Laboratory of
Optics of Palack\'{y} University and Institute of Physics of
Academy of Sciences of the Czech Republic, 17. listopadu 50A, 772
07 Olomouc, Czech Republic}
\email{perinaj@prfnw.upol.cz}
\author{Concita Sibilia}
\author{Mario Bertolotti}
\affiliation{Dipartimento di Energetica, Universit\`{a} ``La
Sapienza'' di Roma, Via A. Scarpa 16, 00161 Roma, Italy}
\author{Michael Scalora}
\affiliation{Charles M. Bowden Research Center, RD\&EC, Redstone
Arsenal, Bldg 7804, Alabama 35898-5000, USA}

\begin{abstract}
Two-mode nonlinear interaction (second-harmonic and
second-subharmonic generation) in a planar waveguide with a small
periodic corrugation at the surface is studied. Scattering of the
interacting fields on the corrugation leads to constructive
interference that enhances the nonlinear process provided that all
the interactions are phase matched. Conditions for the overall
phase matching are found. Compared with a perfectly
quasi-phase-matched waveguide, better values of squeezing as well
as higher intensities are reached under these conditions.
Procedure for finding optimum values of parameters for
squeezed-light generation is described.
\end{abstract}


\maketitle

\section{Introduction}

Since the pioneering work by Armstrong \cite{Armstrong1962} on the
process of second-harmonic generation has occurred,
spatio-temporal properties of the nonlinearly interacting
classical fields have been studied in detail by many authors both
theoretically and experimentally. A new impulse in these studies
has occurred when people understood that this process can give
rise to the fields with nonclassical properties (for a review,
see, e.g. \cite{Perina1991}). Namely light with electric-field
amplitude fluctuations suppressed below the limit given by quantum
mechanics can be generated both in the pump and second-subharmonic
fields. Also light with nonclassical photon-number statistics can
be obtained - pairwise character of photon-number statistics
\cite{Schleich1987,Perina1990} generated in the spontaneous
process of second-subharmonic generation has been observed
\cite{Waks2004}.

It has been shown that the best conditions for squeezed-light
generation in homogeneous nonlinear media occur, provided that the
nonlinear two-mode interaction is perfectly phase matched. Under
these conditions, the principal squeeze variance of the
second-subharmonic field can asymptotically reach zero when the
gain of the nonlinear interaction increases. On the other hand,
the pump-field principal squeeze variance cannot be less than 0.5
\cite{Ou1994}. If large values of the nonlinear phase mismatch are
allowed, this limit can be overcome due to a nonlinear phase
modulation, as suggested in \cite{Li1993,Li1994}. However, the
generated signal is very weak.

The most common method how to compensate for the natural nonlinear
phase mismatch that occurs in commonly used nonlinear materials is
to introduce an additional periodic modulation of the $ \chi^{(2)}
$ susceptibility using periodical poling
\cite{Serkland1995,Serkland1997,Yu2005}. Several methods for the
additional modulation of the local amplitude of this
quasi-phase-matched interaction have been developed
\cite{Huang2006}. These methods allow to reach a spectrally
broad-band two-mode interaction and so femtosecond pumping of the
nonlinear process is possible \cite{Schober2005}.

In order to effectively increase low values of the nonlinear
interaction in real materials, configurations in which a nonlinear
medium is put inside a cavity are usually used to generate
squeezed light (e.g.
\cite{Lawrence2002,Leonhardt1997,Bachor2004}).

In a waveguiding geometry that profits from a strong spatial
confinement of the interacting optical fields in the transverse
plane resulting in high values of the effective nonlinearity,
another method to reach a nonlinear phase mismatch is possible.
One of the nonlinearly interacting fields can be coupled through
its evanescent waves into another field of the same frequency
propagating in a neighbouring waveguide. An exchange of energy
between these two linearly coupled fields introduces a spatial
modulation of the nonlinearly interacting field that can be set to
compensate for the nonlinear phase mismatch \cite{Dong2004}.
Interaction of fields in different waveguides through their
evanescent waves can be used in various configurations that modify
nonclassical properties of optical fields emerging from nonlinear
interactions \cite{PerinaJr2000}.

A waveguiding geometry allows another possibility to tailor the
nonlinear process - a linear periodic corrugation of the waveguide
surface can be introduced as a distributed feedback that scatters
the propagating fields \cite{Joannopoulos1995,Bertolotti2001}.
Under suitable conditions, the scattered contributions interfere
constructively and increase electric-field amplitudes of the
propagating fields. This then results in higher conversion
efficiencies of the nonlinear process
\cite{Haus1998,Pezzetta2001}. Also squeezed-light generation is
supported in this geometry, as discussed in \cite{Tricca2004}
where scattering of the second-harmonic field on the corrugation
has been neglected. A waveguiding geometry with a periodic
corrugation can also be conveniently used for second-harmonic
generation in \v{C}erenkov configuration \cite{Pezzetta2002}. A
general model of squeezed-light generation in nonlinear photonic
structures has been developed in \cite{Sakoda2002}.

In this paper, we show that scattering of the interacting fields
caused by a linear periodic corrugation supports the generation of
squeezed light in both the pump and second-subharmonic fields
provided that an electric-field amplitude of at least one of the
interacting fields is increased inside the waveguide as a
consequence of scattering. We note that an increase of
electric-field amplitudes in the area of periodic corrugation is
small compared with that occurring in layered photonic band-gap
structures (with a deep grating)
\cite{Scalora1997,DAguanno2001,Dumeige2001}.

The generation of squeezed light has been in the center of
attention in quantum optics for more than twenty years. It has
been shown that the squeezed light can be generated also in Kerr
media, nonlinear process of four-wave mixing, and diode lasers
pumped by a sub-Poissonian current (for a review, see
\cite{Bachor2004}). The level of squeezing observable in recent
experiments \cite{Takeno2007,Vahlbruch2007} approaches 10~dB below
the shot-noise level.

The paper is organized as follows. In Sec. II, a quantum model of
the nonlinear interaction including both Heisenberg equations for
operator electric-field amplitudes and model of a generalized
superposition of signal and noise are presented. Conditions for an
efficient squeezed-light generation are derived in Sec. III. A
detailed analysis of the waveguide made of LiNbO$ {}_3 $ is
contained in Sec. IV. Conclusions are drawn in Sec. V. Appendix A
is devoted to mode analysis of an anisotropic planar waveguide.

\section{Model of the nonlinear interaction}

\begin{figure}    
 \resizebox{0.9\hsize}{!}{\includegraphics{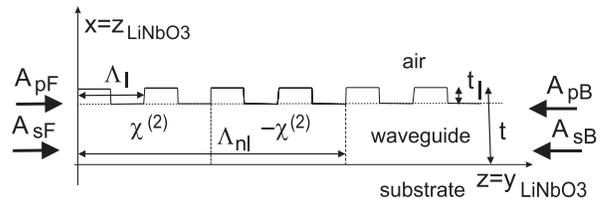}}

 \vspace{2mm}
 \caption{Four optical fields interact in a nonlinear waveguide of thickness $ t $
 and length $ L $ with a periodically-poled
 (period $ \Lambda_{nl} $) $ \chi^{(2)} $ susceptibility;
 $ A_{p_F} $, $ A_{p_B} $, $ A_{s_F} $, and $ A_{s_B} $
 mean forward-propagating pump, backward-propagating
 pump, forward-propagating second-subharmonic, and backward-propagating
 second-subharmonic electric-field amplitudes, respectively.
 A linear corrugation with depth $ t_l $ and period $ \Lambda_l $
 is fabricated on the waveguide upper surface. Profile of the waveguide
 in the $ x-y $ plane is rectangular with width $ \Delta y $ and depth $ t $;
 $ \Delta y \gg t $ is assumed. The waveguide is
 made of LiNbO$ {}_3 $ with the optical axis oriented along the $ x $
 axis.}
 \label{fig1}
\end{figure}

An overall electric-field amplitude $ {\bf E}({\bf r},t) $
describing an optical field in the considered anisotropic
nonlinear waveguide (shown in Fig.~\ref{fig1}) is composed of two
contributions; pump (or fundamental) electric-field amplitude $
{\bf E}_p({\bf r},t) $ at frequency $ \omega $ and
second-subharmonic electric-field amplitude $ {\bf E}_s({\bf r},t)
$ at frequency $ \omega/2 $; i.e. $ {\bf E} = {\bf E}_p + {\bf
E}_s $. We note that in case of second-harmonic generation, the
field with the amplitude $ {\bf E}_p({\bf r},t) $ is called
second-harmonic and that with the amplitude $ {\bf E}_s({\bf r},t)
$ is known as pump. We keep the terminology used for
second-subharmonic generation throughout the paper.
The electric-field amplitude $ {\bf E} $ obeys the wave equation
inside the waveguide with a nonlinear source term
\cite{Snyder1983,Haus1984}:
\begin{equation}     
 \Delta {\bf E} - \nabla( \nabla \cdot {\bf E}) - \mu\epsilon_0 {\bf \epsilon}\cdot
 \frac{\partial^2 {\bf E} }{\partial t^2} = \mu \frac{\partial^2
 {\bf P}_{\rm nl} }{\partial t^2} .
\label{1}
\end{equation}
In Eq.~(\ref{1}) $ \mu $ denotes vacuum permeability, $ \epsilon_0
$ vacuum permittivity, and $ {\bf P}_{\rm nl} $ describes
nonlinear polarization of the medium. The symbol $ \Delta $ stands
for Laplace operator, $ ( \cdot ) $ means a scalar product, and $
\cdot $ denotes tensorial multiplication. Every spectral component
of the element $ {\bf \epsilon}_{ij}({\bf r},\omega) $ of the
relative permittivity tensor in the considered waveguide can be
expressed as follows:
\begin{equation}     
 {\bf \epsilon}_{ij}(x,y,z,\omega) = \bar{\bf \epsilon}_{ij}(x,y,\omega) \left[ 1 +
 \Delta\epsilon_{ij}(x,y,z,\omega) \right] .
\label{2}
\end{equation}
Small variations of permittivity $ {\bf \epsilon} $ described by $
\Delta \epsilon_{ij}(x,y,z,\omega) $ are induced by a periodic
corrugation of the waveguide surface. These variations of the
elements $ \Delta \epsilon_{ij}(x,y,z,\omega) $ along the z axis
can be conveniently decomposed into harmonic waves:
\begin{equation}    
  \Delta\epsilon_{ij}(x,y,z,\omega) = \sum_{q=-\infty}^{\infty}
  {\bf \varepsilon}_{ij,q}(x,y,\omega) \exp\left[ iq \frac{2\pi}{\Lambda_l} z \right],
\label{3}
\end{equation}
where $ {\bf \varepsilon}_{q,ij} $ are coefficients of the
decomposition and $ \Lambda_l $ is a period of the linear
corrugation. Amplitude of the nonlinear polarization $ {\bf
P}_{\rm nl} $ of the medium is determined using tensor $ {\bf d} $
of the second-order nonlinear coefficient:
\begin{equation}    
 {\bf P}_{\rm nl}({\bf r},t) = 2 \epsilon_0 {\bf d}({\bf r})
 \cdot {\bf E}({\bf r},t) {\bf E}({\bf r},t) .
 \label{4}
\end{equation}
Taking into account geometry of our waveguide elements $ {\bf
d}_{ijk}({\bf r}) $ of the nonlinear coefficient can be expressed
as follows:
\begin{equation}     
 {\bf d}_{ijk}(x,y,z) = \sum_{q=-\infty}^{\infty}
  {\bf d}_{ijk,q}(x,y) \exp\left[ iq \frac{2\pi}{\Lambda_{nl}} z \right],
\label{5}
\end{equation}
where $ \Lambda_{nl} $ describes the period of a possible
periodical poling of the nonlinear material.

The electric-field amplitudes of pump ($ {\bf E}_p $) and
second-subharmonic ($ {\bf E}_s $) monochromatic waves can be
expressed in the form:
\begin{eqnarray}    
 {\bf E}_a(x,y,z,t) &=& i \left[ A_{a_F}(z) {\bf e}_a(x,y)
 \exp(i\beta_a z - i\omega_a t) \right. \nonumber \\
 & & \mbox{} + A_{a_B}(z) {\bf e}_a(x,y)
 \exp(-i\beta_a z - i\omega_a t) \nonumber \\
 & & \mbox{} \left. - {\rm H.c.} \right], \hspace{1cm}  a=p,s,
\label{6}
\end{eqnarray}
where the symbols $ {\bf e}_p $ and $ {\bf e}_s $ refer to mode
functions in the transverse plane of the beams. Amplitudes $
A_{p_F} $ and $ A_{s_F} $ [$ A_{p_B} $ and $ A_{s_B} $] describe
forward- [backward-] propagating pump and second-subharmonic
fields and are such that the quantities $ |A_{p_F}|^2 $, $
|A_{p_B}|^2 $, $ |A_{s_F}|^2 $, and $ |A_{s_B}|^2 $ give directly
the number of photons in these fields. Symbol $ \beta_a $ means
propagation constant along the z axis in mode $ a $ whereas $
\omega_a $ stands for frequency of this mode. Symbol $ {\rm H.c.}
$ replaces hermitian conjugated terms. We note that the considered
anisotropic waveguide supports only TM guided modes; TE modes are
not guided and thus do not contribute significantly to nonlinear
interaction.

Mode functions $ {\bf e}_p $ and $ {\bf e}_s $ describe the
transverse profiles of pump and second-subharmonic fields,
respectively, and fulfill the following equations:
\begin{eqnarray}     
  & & \nabla \times \left( \nabla \times \left[{\bf e}_a(x,y)\exp(\pm i \beta_a
  z - i \omega_a t) \right] \right) \nonumber \\
  & & \mbox{} - \frac{\omega_s^2}{c^2}
  \bar{\bf \epsilon}_{ij}(x,y,\omega_a)\cdot \left[{\bf e}_a(x,y)\exp(\pm i \beta_a z
  - i \omega_a t) \right] = 0 , \nonumber \\
  & &  \hspace{3cm} a = p,s.
\label{7}
\end{eqnarray}
We note that mean values of permittivities $ \bar{\bf
\epsilon}_{ij}(x,y,\omega_a) $ are used for the determination of
mode functions. The mode functions $ {\bf e}_p $ and $ {\bf e}_s $
are normalized to describe one photon in a mode inside the
waveguide (see Appendix A).

Substitution of Eqs. (\ref{2}-\ref{6}) into Eq.~(\ref{1}) assuming
$ \left| \frac{\partial^2 A_{a_b}}{\partial z^2} \right| \ll
\left| \beta_a \frac{\partial A_{a_b}}{\partial z} \right| $ for $
a=p,s $ and $ b=F,B $ (analog of the slowly-varying envelope
approximation to spatial evolution) results in the following
equations for amplitudes $ A_{p_F} $, $ A_{p_B} $, $ A_{s_F} $,
and $ A_{s_B} $:
\begin{eqnarray}    
 \frac{dA_{s_F}}{dz} &=& iK_{s}\exp(-i\delta_s z) A_{s_B}
 \nonumber \\
 & & \mbox{} + 4K_{nl,q} \exp(i\delta_{nl,q}z) A_{p_F} A^*_{s_F},
\nonumber\\
\frac{dA_{s_B}}{dz} &=& -iK^*_{s}\exp(i\delta_sz) A_{s_F}
 \nonumber \\
 & & \mbox{} - 4K_{nl,q} \exp(-i\delta_{nl,q}z) A_{p_B} A^*_{s_B},
 \nonumber\\
\frac{dA_{p_F}}{dz} &=& iK_{p} \exp(-i\delta_p z) A_{p_B}
 \nonumber \\
 & & \mbox{} -2K^*_{nl,q}\exp(-i\delta_{nl,q}z)A_{s_F}^2,
 \nonumber\\
\frac{dA_{p_B}}{dz} &=& -iK_{p}^* \exp(i\delta_p z)
 A_{p_F} \nonumber \\
 & & \mbox{} + 2K^*_{nl,q} \exp(i\delta_{nl,q}z) A_{s_B}^2.
\label{8}
\end{eqnarray}
Because the waveguide is made of an anisotropic material, also the
following relation has to hold in order to derive correctly
equations in Eq.~(\ref{8}) (for details, see \cite{Snyder1983}):
\begin{eqnarray}    
 & & \frac{d^2 A_{a_b}(z)}{dz^2} [{\bf e}_a(x,y)]_z \exp(\pm i \beta_a z)
  \; {\bf z} \nonumber\\
 & &  \mbox{} + \frac{d A_{a_b}(z)}{dz} \nabla \left( [{\bf e}_a(x,y)]_z
  \exp(\pm i \beta_a z) \right) \nonumber \\
 & & \mbox{} + \frac{d A_{a_b}(z)}{dz}
  \left( \nabla \cdot \left[ {\bf e}_a(x,y)
  \exp(\pm i \beta_a z) \right] \right) \; {\bf z} \approx 0 ,
  \nonumber\\
 & & \hspace{3cm} a=p,s, \hspace{3mm}  b=F,B,
\label{9}
\end{eqnarray}
and $ {\bf z} $ stands for a unit vector along the $ z $ axis. We
note that Eqs.~(\ref{8}) describe also a nonlinear coupler
composed of two waveguides made of $ \chi^{(2)} $ media whose
modes interact through evanescent waves. Nonclassical properties
of light propagating in this coupler have been studied in
\cite{Perina1996,Korolkova1997}. A scheme that allows to decompose
interactions in this coupler into a sequence of fictitious
interactions has been suggested in \cite{Fiurasek2000}.

Phase mismatches $ \delta_s $, $ \delta_p $, and $ \delta_{nl,q} $
occurring in Eq.~(\ref{8}) are given as follows:
\begin{eqnarray}   
 \delta_a &=& 2\beta_a - \frac{2\pi}{\Lambda_l}, \hspace{3mm} a=p,s
 , \nonumber \\
 \delta_{nl,q} &=& \beta_p - 2 \beta_s + q \frac{2\pi}{\Lambda_{nl}}
 .
\label{10}
\end{eqnarray}
Coefficient $ q $ equals $ \pm 1 $ for a periodically poled
nonlinear material, whereas $ q=0 $ for a material without
periodical poling. Linear coupling constants $ K_p $ and $ K_s $
are determined along the expressions:
\begin{eqnarray}    
 K_a &=& \frac{\omega_a^2}{2c^2\beta_a} \frac{\int dxdy \, {\bf
 \varepsilon}_{1}(x,y,\omega_a) \cdot {\bf e}^*_a(x,y) {\bf e}_a(x,y) }{
 \int dxdy \, |{\bf e}_a(x,y)|^2} ,
 \nonumber \\
 & & a=p,s .
\label{11}
\end{eqnarray}
Similarly, the following expression can be found for nonlinear
coupling constants $ K_{nl,q} $ for $ q=0,\pm 1 $:
\begin{eqnarray}    
 K_{nl,q} &=& \frac{2i\omega_s^2}{c^2\beta_s} \frac{
 \int dxdy \, {\bf d }_q(x,y) \cdot {\bf e}_p(x,y) {\bf
 e}^{*}_s(x,y) {\bf e}^{*}_s(x,y) }{ \int dxdy \, |{\bf e}_s(x,y)|^2}
 \nonumber \\
  &\approx & \frac{2i\omega_p^2}{c^2\beta_p} \frac{
 \int dxdy \, {\bf d }_q(x,y) \cdot {\bf e}_p(x,y) {\bf
 e}^{*}_s(x,y) {\bf e}^{*}_s(x,y) }{ \int dxdy \, |{\bf
 e}_p(x,y)|^2}. \nonumber \\
 & &
\label{12}
\end{eqnarray}
The last approximate equality in Eq.~(\ref{12}) is valid provided
that $ \omega_s/\beta_s \approx \omega_p/\beta_p $ and due to the
normalization of the mode functions $ {\bf e}_s $ and $ {\bf e}_p
$. This approximation assures, that only one nonlinear coupling
constant occurs in Eqs.~(\ref{8}) which is important in quantum
description.

Expressions for linear ($ K_s $, $ K_p $) and nonlinear ($
K_{nl,0} $, $ K_{nl,1} $) coupling constants appropriate for the
considered waveguide and derived from Eqs. (\ref{11}) and
(\ref{12}) can be found in Appendix A in Eqs.
(\ref{A10}-\ref{A12}).

Quantum model of the nonlinear interaction in the considered
waveguide can be formulated changing the classical envelope
electric-field amplitudes $ A_{p_F} $, $ A_{p_B} $, $ A_{s_F} $,
and $ A_{s_B} $ occurring in Eq.~(\ref{6}) into operators denoted
as $ \hat{A}_{p_F} $, $ \hat{A}_{p_B} $, $ \hat{A}_{s_F} $, and $
\hat{A}_{s_B} $, respectively. A quantum analog of the classical
equations written in Eq.~(\ref{8}) can then be derived from the
Heisenberg equations (for details, see
\cite{PerinaJr2004,PerinaJr2005};
\begin{equation}   
 \frac{d\hat{X}}{dz} = - \frac{i}{\hbar} \left[ \hat{G}, \hat{X}
  \right] ;
\label{13}
\end{equation}
considering the following momentum operator $ \hat{G} $:
\begin{eqnarray}    
\hat{G} &=& \left[ \hbar K_s\exp(-i\delta_s z)
 \hat{A}^\dagger_{s_F} \hat{A}_{s_B} \right. \nonumber \\
 & & \mbox{} \left. + \hbar K_p\exp(-i\delta_p z)
 \hat{A}^\dagger_{p_F}
 \hat{A}_{p_B} + {\rm H.c.} \right] \nonumber \\
 & & - \left[ 2i\hbar K_{nl,q} \exp(i\delta_{nl,q}z)
 \hat{A}^{\dagger 2}_{s_F} \hat{A}_{p_F} \right. \nonumber \\
 & & \mbox{} \left. + 2i\hbar K_{nl,q}
 \exp(-i\delta_{nl,q}z) \hat{A}^{\dagger 2}_{s_B} \hat{A}_{p_B}
 + {\rm H.c.} \right] ,
\label{14}
\end{eqnarray}
where $ q=0 $ or $ q=\pm 1$. In Eq.~(\ref{13}), $ \hbar $ means
the reduced Planck constant, $ \hat{X} $ stands for an arbitrary
operator and the symbol $ [,] $ denotes commutator.

The operator quantum equations analogous to those written in
Eq.~(\ref{8}) can be solved using the method of a small operator
correction (denoted as $ \delta \hat{A} $) to a mean value
(denoted as $ A $) in which an operator electric-field amplitude $
\hat{A} $ is decomposed as $ \hat{A} = A + \delta\hat{A} $. This
method provides a set of classical nonlinear equations for the
mean values $ A $ that coincides with the set given in
Eq.~(\ref{8}). The operator electric-field amplitude corrections $
\delta\hat{A} $ fulfill the following linear operator equations:
\begin{eqnarray}     
 \frac{d\delta \hat{A}_{s_F}}{dz} &=& {\cal K}_{s}
  \delta\hat{A}_{s_B}
  + {\cal K}_{F,q}\left[ A_{p_F} \delta \hat{A}^\dagger_{s_F}
  + A^*_{s_F} \delta \hat{A}_{p_F} \right] ,
 \nonumber\\
 \frac{d\delta\hat{A}_{s_B}}{dz} &=& {\cal K}^*_{s}
  \delta\hat{A}_{s_F}
  - {\cal K}_{B,q} \left[ A_{p_B} \delta\hat{A}^*_{s_B}
   + A^*_{s_B} \delta\hat{A}_{p_B} \right],
 \nonumber\\
 \frac{d\delta\hat{A}_{p_F}}{dz} &=& {\cal K}_{p}
  \delta\hat{A}_{p_B}
  -{\cal K}^*_{F,q} A_{s_F} \delta\hat{A}_{s_F},
  \nonumber\\
 \frac{d\delta\hat{A}_{p_B}}{dz} &=& {\cal K}^*_{p}
  \delta\hat{A}_{p_B} + {\cal K}^*_{B,q} A_{s_B}
    \delta\hat{A}_{s_B} .
\label{15}
\end{eqnarray}
The functions $ {\cal K}_s $, $ {\cal K}_p $, and $ {\cal
K}_{nl,q} $ introduced in Eqs. (\ref{15}) are defined as:
\begin{eqnarray}   
 {\cal K}_a &=& iK_a \exp(-i\delta_a z), \hspace{1cm} a=s,p,
 \nonumber \\
 {\cal K}_{F,q} &=& 4K_{nl,q} \exp(i\delta_{nl,q} z), \nonumber \\
 {\cal K}_{B,q} &=& 4K_{nl,q} \exp(-i\delta_{nl,q} z) .
\end{eqnarray}

Solution of the classical nonlinear equations written in Eqs.
(\ref{8}) can only be reached numerically using, e.g., a finite
difference method called BVP \cite{NumericalRecipes}. This method
requires an initial guess of the solution that can be conveniently
obtained when the nonlinear terms in Eqs.~(\ref{8}) are omitted.
Then, the initial solution can be written as follows:
\begin{eqnarray}   
  A_{a_F}^{(0)} &=& \exp\left(-i\frac{\delta_a z}{2}\right) \left[B_a
   \cos(\Delta_a z) + \tilde{B_a} \sin(\Delta_a z) \right],
   \nonumber \\
  A_{a_B}^{(0)} &=& \exp\left(i\frac{\delta_a z}{2}\right) \nonumber \\
   & & \mbox{} \times  \left[B_a
   \left( -\frac{\delta_a}{2K_a}\cos(\Delta_a z) + i\frac{\Delta_a}{K_a}
   \sin(\Delta_a z) \right) \right.
   \nonumber \\
   & & \mbox{} \left. + \tilde{B_a} \left( -\frac{\delta_a}{2K_a}\sin(\Delta_a z)
   - i\frac{\Delta_a}{K_a} \cos(\Delta_a z) \right) \right],
   \nonumber \\
   & & a=s,p,
\label{17}
\end{eqnarray}
and
\begin{equation}  
 \Delta_a = \sqrt{\frac{\delta_a^2}{4} - |K_a|^2}, \hspace{1cm}
   a=s,p.
\end{equation}
In Eqs.~(\ref{17}), constants $ B_{p} $, $ \tilde{B_{p}} $, $
B_{s} $, and $ \tilde{B_{s}} $ are set according to the boundary
conditions at both sides of the waveguide.

We note that any solution of the nonlinear equations in Eqs.
(\ref{8}) obeys the following relation useful in a numerical
computation:
\begin{equation}     
 \frac{d}{dz} \left( |A_{s_F}|^2 + 2 |A_{p_F}|^2
   - |A_{s_B}|^2 - 2 |A_{p_B}|^2 \right) = 0.
\label{19}
\end{equation}

The solution of the system of linear operator equations in Eqs.
(\ref{15}) for the operator electric-field amplitude corrections $
\delta\hat{A} $ can be found numerically and put into the
following matrix form:
\begin{equation}     
 \pmatrix{\delta \hat{\cal A}_{F,\rm out} \cr \delta \hat{\cal
 A}_{B,\rm in}}
 =  \pmatrix{ {\cal U}_{FF} & {\cal U}_{FB} \cr
 {\cal U}_{BF} & {\cal U}_{BB} }
 \pmatrix{\delta \hat{\cal A}_{F,\rm in} \cr
  \delta \hat{\cal A}_{B,\rm out}},
 \label{20}
\end{equation}
where
\begin{eqnarray}      
 \delta \hat{\cal A}_{F,\rm in} = \pmatrix{\delta \hat{A}_{s_F}(0)
 \cr \delta \hat{A}^\dagger_{s_F}(0) \cr \delta \hat{A}_{p_F}(0)
 \cr \delta \hat{A}^\dagger_{p_F}(0)} , \hspace{0.2cm} \delta
 \hat{\cal A}_{F,\rm out} = \pmatrix{ \delta \hat{A}_{s_F}(L) \cr
 \delta \hat{A}^\dagger_{s_F}(L) \cr \delta \hat{A}_{p_F}(L) \cr
 \delta \hat{A}^\dagger_{p_F}(L)} , \nonumber \\
 \delta \hat{\cal A}_{B,\rm in} = \pmatrix{ \delta \hat{A}_{s_B}(L)
 \cr \delta \hat{A}^\dagger_{s_B}(L) \cr \delta \hat{A}_{p_B}(L)
 \cr \delta \hat{A}^\dagger_{p_B}(L) }, \hspace{0.2cm} \delta
 \hat{\cal A}_{B,\rm out} = \pmatrix{ \delta \hat{A}_{s_B}(0) \cr
 \delta \hat{A}^\dagger_{s_B}(0) \cr \delta \hat{A}_{p_B}(0) \cr
 \delta \hat{A}^\dagger_{p_B}(0) },
\label{21}
\end{eqnarray}
and $ L $ means the length of the waveguide. Matrices $ {\cal
U}_{FF} $, $ {\cal U}_{FB} $, $ {\cal U}_{BF} $, and $ {\cal
U}_{BB} $ are determined using the numerical solution of Eqs.
(\ref{15}).

The following input-output relations among the operator amplitude
corrections $ \delta\hat{A} $,
\begin{eqnarray}     
 \pmatrix{\delta \hat{A}_{F,\rm out} \cr
 \delta \hat{A}_{B,\rm out}}  &=&
 \pmatrix{ {\cal U}_{FF}-{\cal U}_{FB} {\cal U}^{-1}_{BB}
 {\cal U}_{BF} & {\cal U}_{FB} {\cal U}^{-1}_{BB} \cr
 -{\cal U}^{-1}_{BB}{\cal U}_{BF} & {\cal U}^{-1}_{BB}}
 \nonumber \\
 & & \mbox{} \times
  \pmatrix{\delta \hat{\cal A}_{F,\rm in} \cr \delta \hat{\cal
  A}_{B,\rm in}}
  \label{22} \\
  &=&  {\cal U} \pmatrix{\delta \hat{\cal A}_{F,\rm in} \cr \delta \hat{\cal
  A}_{B,\rm in}},
  \label{23}
\end{eqnarray}
are found solving Eqs. (\ref{21}) with respect to vectors $ \delta
\hat{\cal A}_{F,\rm out} $ and $ \delta \hat{\cal A}_{B,\rm out}
$. The output operator electric-field amplitude corrections
contained in vectors $ \delta \hat{\cal A}_{F,\rm out} $ and $
\delta \hat{\cal A}_{B,\rm out} $ obey boson commutation relations
provided that the input operator electric-field amplitude
corrections given in vectors $ \delta \hat{\cal A}_{F,\rm in} $
and $ \delta \hat{\cal A}_{B,\rm in} $ are ruled by boson
commutation relations. We note that also certain commutation
relations among the operator electric-field amplitude corrections
in vectors $ \delta \hat{\cal A}_{F,\rm out} $ and $ \delta
\hat{\cal A}_{B,\rm in} $ can be derived (for details, see
\cite{PerinaJr2005}).

We restrict our considerations to states of optical fields that
can be described using the generalized superposition of signal and
noise \cite{Perina1991}. Thus coherent states, squeezed states as
well as noise can be considered. Parameters $ B_j $, $ C_j $, $
D_{jk} $, and $ \bar{D}_{jk} $ defined below are sufficient for
the description of any state of a two-mode optical field in this
approximation \cite{PerinaJr2000}:
\begin{eqnarray}   
 B_j &=& \langle \Delta\hat{A}^\dagger_j \Delta\hat{A}_j \rangle ,
  \nonumber \\
 C_j &=& \langle (\Delta\hat{A}_j)^2 \rangle ,
  \nonumber \\
 D_{jk} &=& \langle \Delta\hat{A}_j \Delta\hat{A}_k \rangle ,
  \hspace{1cm} j \neq k,  \nonumber \\
 \bar{D}_{jk} &=& -\langle \Delta\hat{A}^\dagger_j \Delta\hat{A}_k
  \rangle , \hspace{1cm} j \neq k;
\label{24}
\end{eqnarray}
$ \Delta\hat{A}_j = \hat{A}_j - \langle\hat{A}_j\rangle $; symbol
$ \langle \;\; \rangle $ denotes the quantum statistical mean
value. Expressions for the coefficients $ B_j $, $ C_j $, $ D_{jk}
$, and $ \bar{D}_{jk} $ appropriate for outgoing fields can be
derived \cite{PerinaJr2005} using elements of the matrix $ \cal U
$ given in Eq. (\ref{23}) and incident values of the coefficients
$ B_{j,\rm in,{\cal A}} $ and $ C_{j,\rm in,{\cal A}} $ related to
anti-normal ordering of field operators (for details, see
\cite{PerinaJr2000}):
\begin{eqnarray}   
 B_{j,\rm in,{\cal A}} &=& \cosh^2(r_{j}) + n_{ch,j} ,
  \nonumber \\
 C_{j,\rm in,{\cal A}} &=& \frac{1}{2} \exp (i\vartheta_j)
  \sinh (2r_j) .
\label{25}
\end{eqnarray}
In Eq.~(\ref{25}), $ r_j $ stands for a squeeze parameter of the
incident $ j $-th field, $ \vartheta_j $ means a squeeze phase,
and $ n_{ch,j} $ stands for a mean number of incident chaotic
photons. Coefficients $ D_{jk,\rm in,{\cal A}} $ and $
\bar{D}_{jk,\rm in,{\cal A}} $ for an incident field are
considered to be zero, i.e. the incident fields are assumed to be
statistically independent.

The maximum attainable value of squeezing of electric-field
amplitude fluctuations is given by the value of a principal
squeeze variance $ \lambda $ \cite{Perina1991}. Both single-mode
principal squeeze variances $ \lambda_j $ and compound-mode
principal squeeze variances $ \lambda_{ij} $ (characterizing an
overall field composed of two other fields) can be determined in
terms of the coefficients $ B_j $, $ C_j $, $ D_{jk} $, and $
\bar{D}_{jk} $ given in Eq.~(\ref{24}) (for details, see,
\cite{Perina1991,PerinaJr2000}):
\begin{eqnarray}     
\lambda_j &=& 1+ 2[B_j-|C_j|], \\
\lambda_{jk} &=& 2\left[1+ B_j+B_k -
  2\Re(\bar{D}_{jk}) \right. \nonumber \\
  & & \mbox{} \left. - | C_j+C_k+2D_{jk}| \right] ;
\end{eqnarray}
symbol $ \Re $ means the real part of an expression. Values of the
principal squeeze variance $ \lambda_j $ ($ \lambda_{jk} $) less
than one (two) indicate squeezing in a single-mode (compound-mode)
case.

\section{Suitable conditions for squeezed-light generation}

It occurs that there exist two conditions for an efficient
squeezed-light generation. The first condition comes from the
requirement that the nonlinear interaction should be phase-matched
along the whole waveguide, whereas the second one gives optimum
conditions for the enhancement of electric-field amplitudes of the
interacting optical fields inside the waveguide.

\subsection{Overall phase-matching of the nonlinearly interacting fields}

Conditions for an optimum phase-matching of the interacting fields
can be revealed, when we write the differential equation for the
number $ N_{a} $ of photons in field $ a $; $ N_a = A^*_a A_a $.
Using Eqs.~(\ref{8}) we arrive at the following differential
equations:
\begin{eqnarray}   
 \frac{dN_{s_F}}{dz} &=& -2 \Im \left\{K_{s}\exp(-i\delta_s z)
 A^*_{s_F}A_{s_B} \right\}
 \nonumber \\
 & & \mbox{} + 8\Re \left\{K_{nl,q} \exp(i\delta_{nl,q}z) A^{*2}_{s_F} A_{p_F}
  \right\},  \nonumber \\
 \frac{dN_{s_B}}{dz} &=& -2 \Im \left\{K_{s}\exp(-i\delta_s z)
 A^*_{s_F}A_{s_B} \right\}
 \nonumber \\
 & & \mbox{} - 8\Re \left\{K_{nl,q} \exp(-i\delta_{nl,q}z) A^{*2}_{s_B} A_{p_B}
  \right\}, \nonumber \\
 \frac{dN_{p_F}}{dz} &=& -2 \Im \left\{K_{p}\exp(-i\delta_p z)
 A^*_{p_F}A_{p_B} \right\}
 \nonumber \\
 & & \mbox{} - 4\Re \left\{K_{nl,q} \exp(i\delta_{nl,q}z) A^{*2}_{s_F} A_{p_F}
  \right\}, \nonumber \\
 \frac{dN_{p_B}}{dz} &=& -2 \Im \left\{K_{p}\exp(-i\delta_p z)
 A^*_{p_F}A_{p_B} \right\}
 \nonumber \\
 & & \mbox{} + 4\Re \left\{K_{nl,q} \exp(-i\delta_{nl,q}z) A^{*2}_{s_B} A_{p_B}
  \right\};
 \label{28}
\end{eqnarray}
symbol $ \Im $ denotes the imaginary part of an expression. We
note that the first terms on the right-hand side of the first and
the second (as well as the third and the fourth) equations in
Eqs.~(\ref{28}) have the same sign because of counter-propagation
of the fields [see also Eq.~(\ref{19})]. The nonlinear interaction
described by the second terms on the right-hand sides of
Eqs.~(\ref{28}) is weak and so we can judge the contribution of
these terms using a perturbation approach. In the first step we
neglect the nonlinear terms in Eqs.~(\ref{8}) and solve
Eqs.~(\ref{8}) for field amplitudes $ A_{s_F}^{(0)} $, $
A_{s_B}^{(0)} $, $ A_{p_F}^{(0)} $, and $ A_{p_B}^{(0)} $. Then we
insert this solution into the nonlinear terms in Eqs.~(\ref{28})
and find this way optimum conditions that maximize contributions
of these terms. The solution for amplitudes $ A_{s_F}^{(0)} $, $
A_{s_B}^{(0)} $, $ A_{p_F}^{(0)} $, and $ A_{p_B}^{(0)} $
coincides with that written in Eqs.~(\ref{17}) as an initial guess
for the numerical solution and we rewrite it into the following
suitable form:
\begin{eqnarray}     
  A_{a_F}^{(0)}(z) &=& \exp\left(-i\frac{\delta_a z}{2}\right) \nonumber \\
   & & \mbox{} \times \left[{\cal B}_{a_F}^+ \exp(i\Delta_a z) + {\cal B}_{a_F}^-
   \exp(-i\Delta_a z) \right],
   \nonumber \\
   & & {\cal B}_{a_F}^+ = \frac{B_a - i\tilde{B}_a}{2} , \nonumber \\
   & & {\cal B}_{a_F}^- = \frac{B_a + i\tilde{B}_a}{2} , \nonumber \\
  A_{a_B}^{(0)}(z) &=& \exp\left(i\frac{\delta_a z}{2}\right) \nonumber \\
   & & \mbox{} \times \left[ {\cal B}_{a_B}^+ \exp(i\Delta_a z) + {\cal B}_{a_B}^- \exp(-i\Delta_a
   z)\right] , \nonumber \\
   & & {\cal B}_{a_B}^+ = \frac{-\delta_a + 2\Delta_a}{4K_a} (B_a - i\tilde{B}_a) ,
     \nonumber \\
   & & {\cal B}_{a_B}^- = \frac{-\delta_a - 2\Delta_a}{4K_a} (B_a +
   i\tilde{B}_a) , a=p,s. \nonumber \\
   & &
\label{29}
\end{eqnarray}
Provided that a linear corrugation is missing in field $ a $ the
solution for amplitudes $ A_{a_F}^{(0)} $ and $ A_{a_B}^{(0)} $
can be obtained from the expressions in Eqs.~(\ref{29}) using a
sequence of two limits; $ \delta_a \rightarrow 0 $, $ K_a
\rightarrow 0 $:
\begin{eqnarray}     
  A_{a_F}^{(0)}(z) &=& B_a , \nonumber \\
  A_{a_B}^{(0)}(z) &=& \tilde{B}_a.
\label{30}
\end{eqnarray}

The nonlinear interaction between the forward-propagating fields
is described in Eqs.~(\ref{28}) in our perturbation approach by
the term
\begin{equation}    
 \Re \left\{K_{nl,q} \exp(i\delta_{nl,q}z) {A^{(0)}_{s_F}}^{*2}
 A_{p_F}^{(0)} \right\}
\end{equation}
that, after substituting the expressions for amplitudes $
A_{s_F}^{(0)} $ and $ A_{p_F}^{(0)} $ from Eqs.~(\ref{29}), splits
into the following eight terms:
\begin{eqnarray}    
 & & \hspace{-0.5cm}
  \Re \biggl\{ K_{nl,q} {{\cal B}_{s_F}^{\pm}}^* {{\cal B}_{s_F}^{\pm}}^*
  {\cal B}_{p_F}^{\pm} \nonumber \\
 & &  \hspace{-0.2cm} \mbox{} \times   \exp(i[\delta_{nl,q} + \delta_s
  -\delta_p/2 \mp \Delta_s \mp \Delta_s \pm \Delta_p ]z) \biggr\} .
\end{eqnarray}
The nonlinear interaction is efficient under the condition that
one of these terms does not oscillate along the $ z $ axis. This
gives us eight possible conditions that combine nonlinear
phase-mismatch $ \delta_{nl,q} $ and parameters of the corrugation
$ \delta_s $, $\delta_p $, $ K_s $, and $ K_p $:
\begin{equation}   
 \delta_{nl,q} + \delta_s -\delta_p/2
 \mp \Delta_s \mp \Delta_s \pm \Delta_p = 0.
\label{33}
\end{equation}
It depends on a given waveguide and initial conditions which out
of these eight conditions leads to an efficient nonlinear
interaction. We note that the conditions in Eq.~(\ref{33}) are
valid also for the nonlinear interaction between the
backward-propagating fields characterized by the term $ \Re
\left\{K_{nl,q} \exp(-i\delta_{nl,q}z) {A^{(0)}_{s_B}}^{*2}
 A_{p_B}^{(0)} \right\} $ in Eqs.~(\ref{28}).

We consider two special cases in which a periodic corrugation is
present either in the pump or the second-subharmonic field.
Assuming the corrugation in the pump field the conditions in
Eq.~(\ref{33}) get the form:
\begin{equation}   
 \delta_{nl,q} -\delta_p/2 \pm \Delta_p = 0.
\label{34}
\end{equation}
Sign - (+) is suitable for $ \delta_{nl,q} > 0 $ ($ \delta_{nl,q}
< 0 $) when we solve Eq.~(\ref{34}) for $ \delta_p $:
\begin{equation}  
 \delta_p = \delta_{nl,q} + \frac{|K_p|^2}{\delta_{nl,q}};
\label{35}
\end{equation}
i.e. $ \delta_{nl,q} $ and $ \delta_p $ have the same sign. The
expression in Eq.~(\ref{35}) then determines the period $
\Lambda_l $ of linear corrugation:
\begin{equation}   
 \Lambda_l = \frac{\pi}{\beta_p - \delta_{nl,q}/2 - |K_p|^2/
 (2\delta_{nl,q})} .
\end{equation}

On the other hand, the conditions
\begin{equation}   
 \delta_{nl,q} +\delta_s/2 \mp \Delta_s \mp \Delta_s = 0
\label{37}
\end{equation}
are suitable for the periodic corrugation in the
second-subharmonic field. When $ \delta_{nl,q} > 0 $ ($
\delta_{nl,q} < 0 $) signs - (+) in Eq.~(\ref{37}) are appropriate
and we have:
\begin{equation}  
 \delta_s = - \frac{\delta_{nl,q}}{2} - \frac{2|K_s|^2}{\delta_{nl,q}};
\label{38}
\end{equation}
i.e. $ \delta_{nl,q} $ and $ \delta_s $ have the opposite sign.
The period $ \Lambda_l $ of linear corrugation is then given as
\begin{equation}   
 \Lambda_l = \frac{\pi}{\beta_s + \delta_{nl,q}/4 + |K_p|^2/
 \delta_{nl,q}} .
\label{39}
\end{equation}

\subsection{Enhancement of amplitudes of the interacting fields}

The greatest increase of electric-field amplitudes inside the
waveguide occurs under the condition of transparency for the
incoming field \cite{Haus1998}. We note that this property also
occurs in layered structures with band-gaps and transmission peaks
where the field is well localized inside the structure provided
that it lies in a transmission peak \cite{Scalora1997}.

A transmission peak of the waveguide can be found from the
condition that a backward-propagating field is zero both at the
beginning and at the end of the waveguide, because it is not
initially seeded. These requirements are fulfilled by the solution
in Eqs.~(\ref{17}) under the following conditions:
\begin{eqnarray}    
 & & \Delta_a = \frac{m\pi}{L}, \hspace{2mm} m = 1,2,\ldots ; \nonumber \\
 & & B_a = A_{a_F}^{(0)}(0); \hspace{3mm} \tilde{B}_a =
 \frac{i\delta_a}{2\Delta_a} B_a, \hspace{2mm} a=p,s.
\label{40}
\end{eqnarray}
A natural number $ m $ counts areas of transmission.

The linear phase mismatch $ \delta_a $ is determined from the
first equation in Eqs.~(\ref{40}) in the following form
\begin{equation}   
 \delta_a = \pm 2 \sqrt{ \left(\frac{m\pi}{L}\right)^2 + |K_a|^2
 }, \hspace{2mm} a=p,s,
\label{41}
\end{equation}
and the corresponding period $ \Lambda_l $ of linear corrugation
is given as:
\begin{equation}    
 \Lambda_l = \frac{\pi}{\beta_a \pm
 \sqrt{ (m\pi/L)^2 + |K_a|^2}}.
\end{equation}

The conditions in Eqs.~(\ref{35}) and (\ref{41}) for a periodic
corrugation in the pump field can be combined together to provide
the following formula for the coupling constant $ K_p $:
\begin{equation}  
 |K_p|^2 = \delta_{nl,q}^2 \left( 1 \pm
 \frac{2m\pi}{|\delta_{nl,q}|L} \right).
\label{43}
\end{equation}
The linear phase mismatch $ \delta_p $ is then determined along
the formula in Eq.~(\ref{41}) such that the signs of $
\delta_{nl,q} $ and $ \delta_p $ are the same.

Similarly the conditions for a periodic corrugation in the
second-subharmonic field written in Eqs.~(\ref{38}) and (\ref{41})
and considered together lead to an expression for the coupling
constant $ K_s $:
\begin{equation}  
 |K_s|^2 = \frac{\delta_{nl,q}^2}{4} \left( 1 \pm
 \frac{4m\pi}{|\delta_{nl,q}|L} \right).
\end{equation}
The sign of linear phase mismatch $ \delta_s $ determined from
Eq.~(\ref{41}) is opposite to that of $ \delta_{nl,q} $.

\section{Squeezed-light generation - numerical analysis}

The discussion of squeezed-light generation is decomposed into
three parts. In the first one, we pay a detailed attention to
attainable characteristics of the considered waveguide. The second
part is devoted to second-subharmonic generation, i.e. an incident
strong pump field is assumed. In the third part, a strong incident
second-subharmonic field is assumed, i.e. second-harmonic
generation is studied.

\subsection{Characteristic parameters of the waveguide}

We consider a waveguide made of LiNbO$ {}_3 $ with length $ L = 1
\times 10^{-3} $~m and width $ \Delta y = 1 \times 10^{-5} $~m
pumped at the wavelength $ \lambda_p = 0.534 \times 10^{-6} $~m ($
\lambda_s = 1.064 \times 10^{-6} $~m). More details are contained
in Appendix~A. We require a single-mode operation at both the
pump- and second-subharmonic-field frequencies that can be
achieved for values of the thickness $ t $ of the waveguide in the
range $ t\in(0.403, 0.544) \times 10^{-6} $~m. We also neglect
losses in the waveguide that may be caused both by absorption
inside the waveguide and scattering of light on the corrugation
that does not propagate into guided modes. In practise, material
absorption is weak at the used wavelengths. Scattering of light on
imperfections of both the guided structure and periodic
corrugation determines losses in a real waveguide and leads to
degradation of squeezing. We also assume that intensity of the
pump field is such that self-phase modulation due to $ \chi^{3} $
nonlinearity does not occur.

\begin{figure}    
 {\raisebox{3.3 cm}{a)} \hspace{5mm}
 \resizebox{0.65\hsize}{!}{\includegraphics{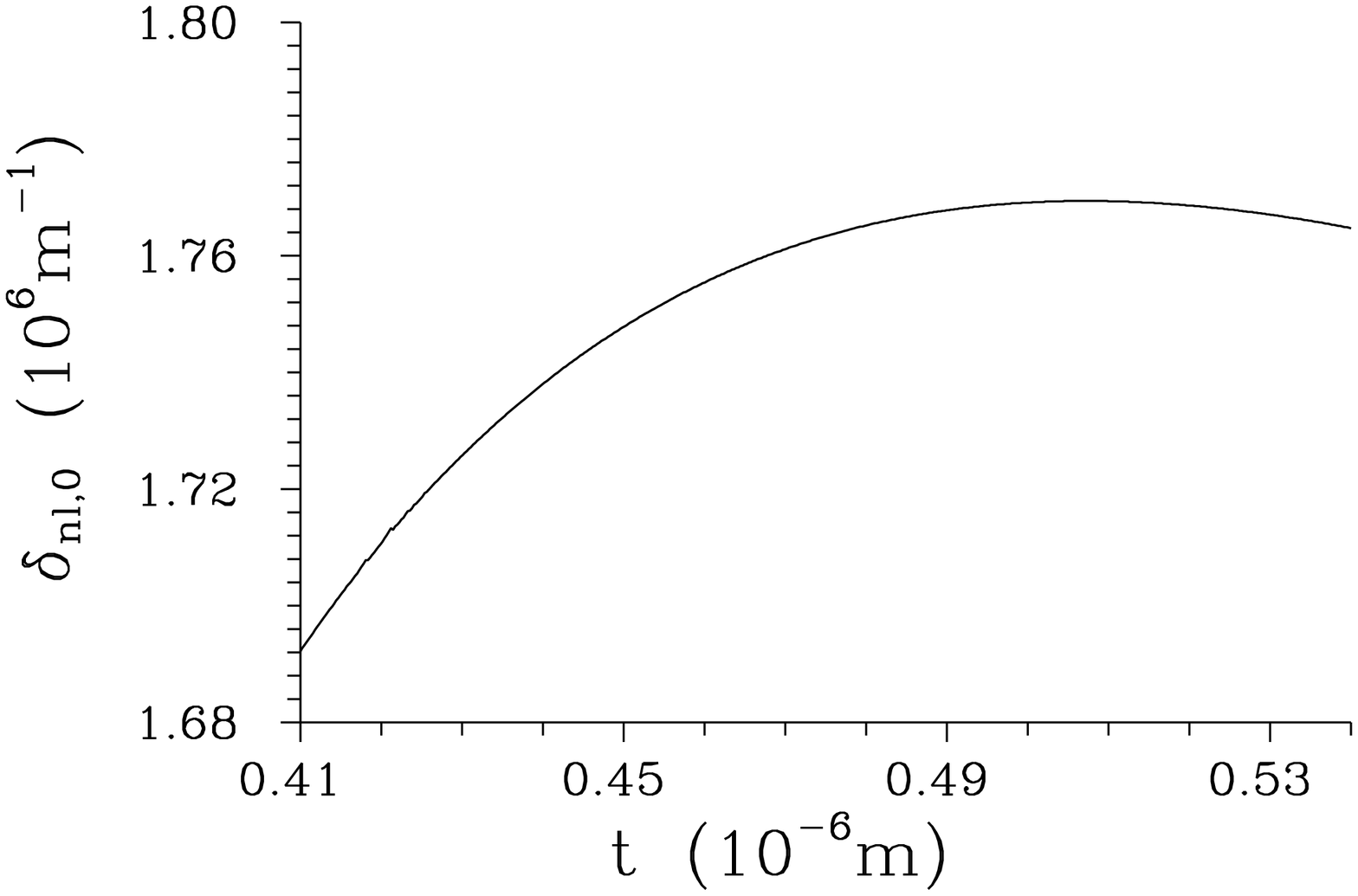}}

 \vspace{0mm}
 \raisebox{3.3 cm}{b)} \hspace{9mm}
 \resizebox{0.6\hsize}{!}{\includegraphics{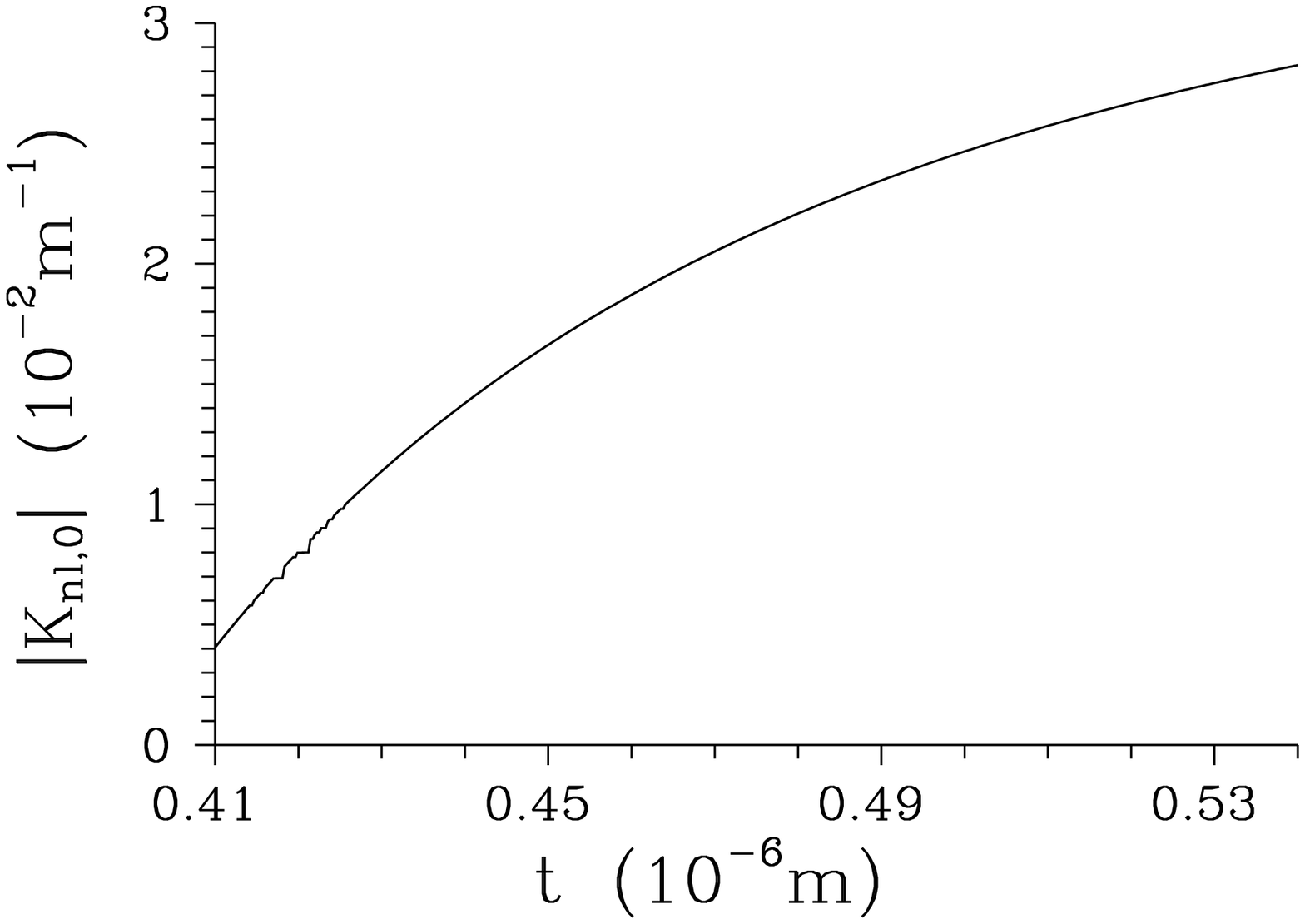}} }
 \vspace{0mm}

 \caption{Natural nonlinear phase mismatch $ \delta_{nl,0} $ (a) and
 absolute value of the nonlinear coupling constant $ K_{nl,0} $ (b)
 as they depend on the thickness
 $ t $ of the waveguide are shown in the region of single-mode operation;
 constant $ K_{nl,0} $ is determined for amplitudes that correspond to one pump
 and one second-subharmonic photon inside the waveguide.}
\label{fig2}
\end{figure}

Values of the natural nonlinear phase mismatch $ \delta_{nl,0} $
are high (around $ 1.7 \times 10^6 $~m$ {}^{-1} $) in the region
of single-mode operation and depend on the thickness $ t $ of the
waveguide (see Fig.~\ref{fig2}a). An additional periodical poling
has to be introduced to compensate, at least partially, for this
mismatch and allow an efficient nonlinear process
\cite{Tricca2004}. Values of the nonlinear coupling constant $
K_{nl,0} $ depicted in Fig.~\ref{fig2}b increase with the
increasing values of the thickness $ t $ because of the thicker
the waveguide the more the modes are localized inside the
waveguide and so the greater the overlap of the nonlinearly
interacting electric-field amplitudes. Attainable values of the
linear coupling constants $ K_p $ and $ K_s $ for different values
of thickness $ t $ and depth $ t_l $ of corrugation can be
obtained from Fig.~{\ref3}. Values of the coupling constants $ K_p
$ and $ K_s $ for a given thickness $ t $ increase monotonously
with the increasing values of the depth $ t_l $ of corrugation.
For small values of the thickness $ t $, values of the coupling
constant $ K_s $ are small because the waveguide is thin for the
second-subharmonic field and so a considerable part of the field
is outside the waveguide and cannot be scattered by the
corrugation.
\begin{figure}[h]    
 {\raisebox{0 cm}{a)} \hspace{5mm}
 \resizebox{1.\hsize}{!}{\includegraphics{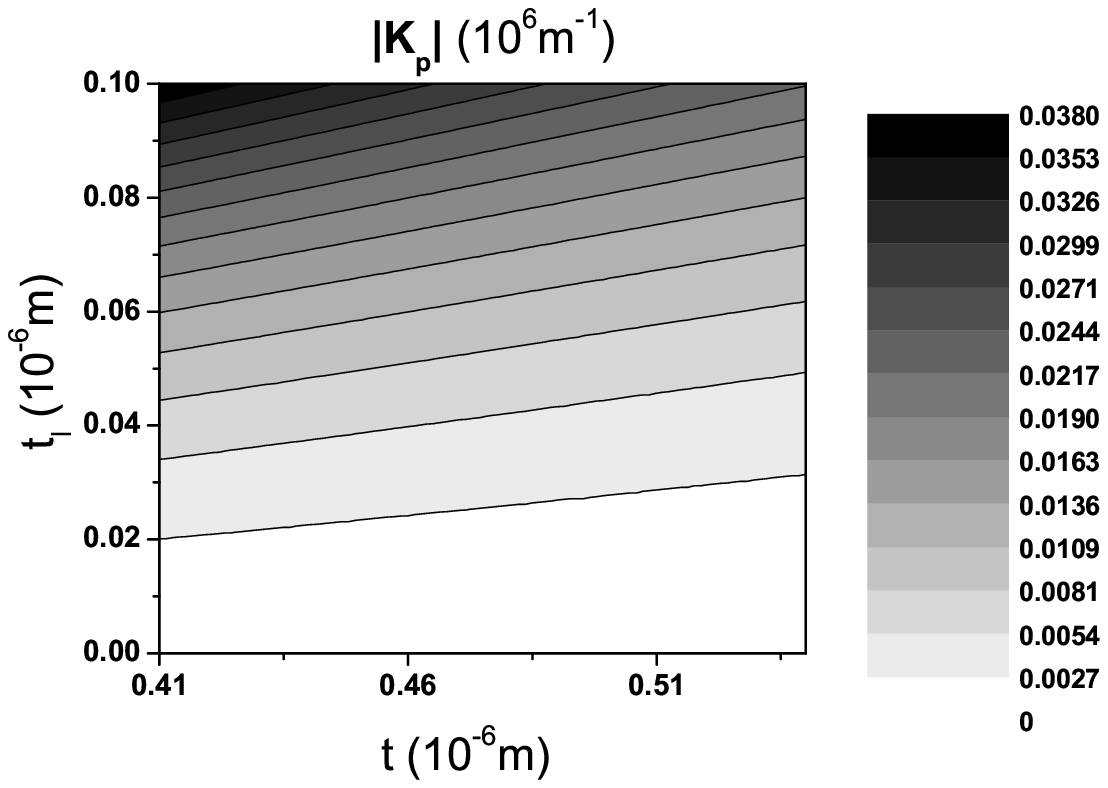}}

 \vspace{0mm}
 \raisebox{0 cm}{b)} \hspace{5mm}
 \resizebox{1.\hsize}{!}{\includegraphics{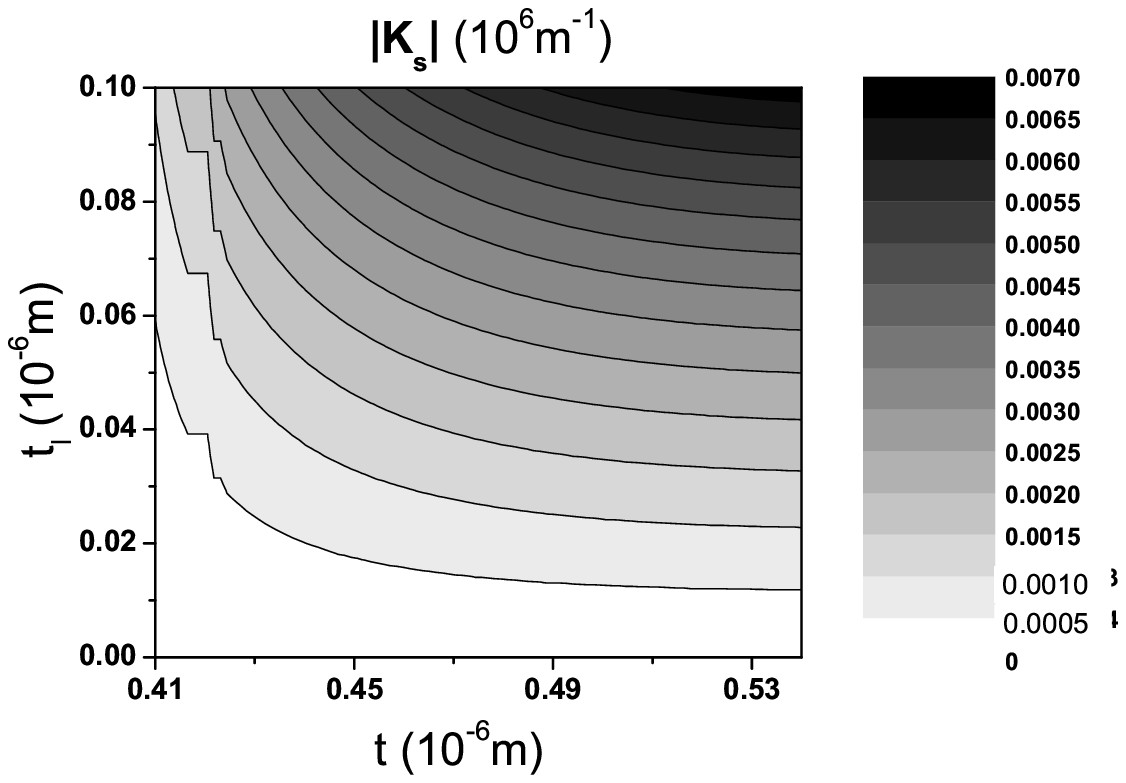}}}
 \vspace{0mm}

 \caption{Contour plot of the absolute value of pump-field
  [second-subharmonic-field] linear coupling constant $ K_p $
  [$ K_s $] as it depends on thickness $ t $ of the waveguide and depth $ t_l $
  of linear corrugation is shown in (a) [(b)] in the region of single-mode operation.}
\label{fig3}
\end{figure}

Amplitudes $ |A_{a_b}| $ of the dimensionless incident strong
electric-field amplitudes are determined from the incident power $
P_{a_b} $ along the relation:
\begin{equation}    
 |A_{a_b}| = \sqrt{\frac{P_{a_b}L\beta_a}{\hbar\omega_a^2}},
 \hspace{3mm} a=p,s, \hspace{3mm}  b=F,B.
\end{equation}
On the other hand, the power $ P_{a_b}^{\rm out} $ of an outgoing
field is given as follows:
\begin{eqnarray}    
 P_{a_b}^{\rm out} &=& \frac{\hbar\omega_a^2}{\beta_a L}
 |A_{a_b}|^2 \nonumber \\
 &=& \frac{\hbar\omega_a^2}{\beta_a L} N_{a_b}, \hspace{3mm}
 a=p,s, \hspace{3mm} b=F,B;
\end{eqnarray}
$ N_{a_b} $ denotes the number of photons leaving the waveguide.

New dimensionless parameters are convenient for the discussion of
behavior of the waveguide;
\begin{eqnarray}    
 & & z^r = z/L , \nonumber \\
 & & \Lambda_l^r = \Lambda_l /L ,
  \Lambda_{nl,q}^r=\Lambda_{nl,q}/L, \nonumber \\
 & & \beta_a^r = L \beta_a, \delta_a^r = L \delta_a, K_a^r = LK_a,
  \hspace{3mm} a=p,s, \nonumber \\
 & & \delta_{nl,q}^r = L \delta_{nl,q}, K_{nl,q}^r = LK_{nl,q}.
\end{eqnarray}
Applying these parameters the waveguide extends from $ z^r = 0 $
to $ z^r = 1 $. The dimensionless parameters enable to understand
the behavior of the waveguide as it depends on the length $ L $
using the graphs and discussion bellow.

\subsection{Second-subharmonic generation}

\begin{figure}    
 \resizebox{0.6\hsize}{!}{\includegraphics{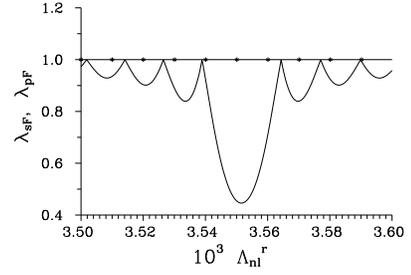}}
 \vspace{0mm}

 \caption{Principal squeeze variances $ \lambda_{s_F} $ (solid
  curve) and $ \lambda_{p_F} $ (solid curve with $ \ast $) as functions of
  the period $ \Lambda_{nl}^r $ of poling for
  second-subharmonic generation;
  to understand the behavior with respect to the length $ L $, the
  period $ \Lambda_{nl} $ should be decomposed as $ \Lambda_{nl}^0
  + \Delta \Lambda_{nl} $, then $ \delta_{nl,q}^r \approx - 2\pi q
  \Delta\Lambda_{nl}^r /{\Lambda_{nl}^0}^2 $, $ \Lambda_{nl}^0 =
  2\pi q/ (2\beta_s-\beta_p) $, $ \Delta\Lambda_{nl}^r =
   L\Delta\Lambda_{nl} $;
  $ t = 5 \times 10^{-7} $~m, $ K_p^r = 0 $, $ K_s^r =0 $,
  $ P_{p_F} = 2 $~W, $ P_{s_F} = 1 \times 10^{-10} $~W (a negligible
  seeding to substitute the spontaneous process in the classical equations),
  $ \arg(A_{p_F}) = 0 $,
  $ \arg(A_{s_F}) = 0 $, $ A_{p_B} = 0 $, $ A_{s_B} = 0 $;
  $ \Lambda_{nl}^0 \approx 3.552 \times 10^{-6} $~m.}
\label{fig4}
\end{figure}

As a reference for the efficiency of squeezed-light generation we
consider the waveguide with periodical poling and assume that it
is pumped by the power of 2~W. The nonlinear interaction is
perfectly phase matched for the period of poling $ \Lambda_{nl}^r
\approx 3.552 \times 10^{-3} $ where we have for the principal
squeeze variances $ \lambda_{s_F} \approx 0.45 $ and $
\lambda_{p_F} \approx 1 $ (see Fig.~\ref{fig4}). The more distant
the value of $ \Lambda_{nl}^r $ from the above-mentioned optimum
value is, the larger the nonlinear phase mismatch and the larger
the value of the principal squeeze variance $ \lambda_{s_F}$.

Considering a periodic corrugation in the pump field with such
parameters that enhancement of the pump field inside the waveguide
occurs, better values of squeezing in the second-subharmonic field
can be reached. However, nonzero values of the nonlinear phase
mismatch $ \delta_{nl,1} $ are important to reach better squeezing
because they have to compensate periodic spatial oscillations
caused by the corrugation (see \cite{Schober2005a}). A perfect
phase matching of all the processes occurring in the waveguide can
be reached this way [see Eq.~(\ref{34})].

\begin{figure}    
 {\raisebox{0 cm}{a)} \hspace{5mm}
 \resizebox{1.\hsize}{!}{\includegraphics{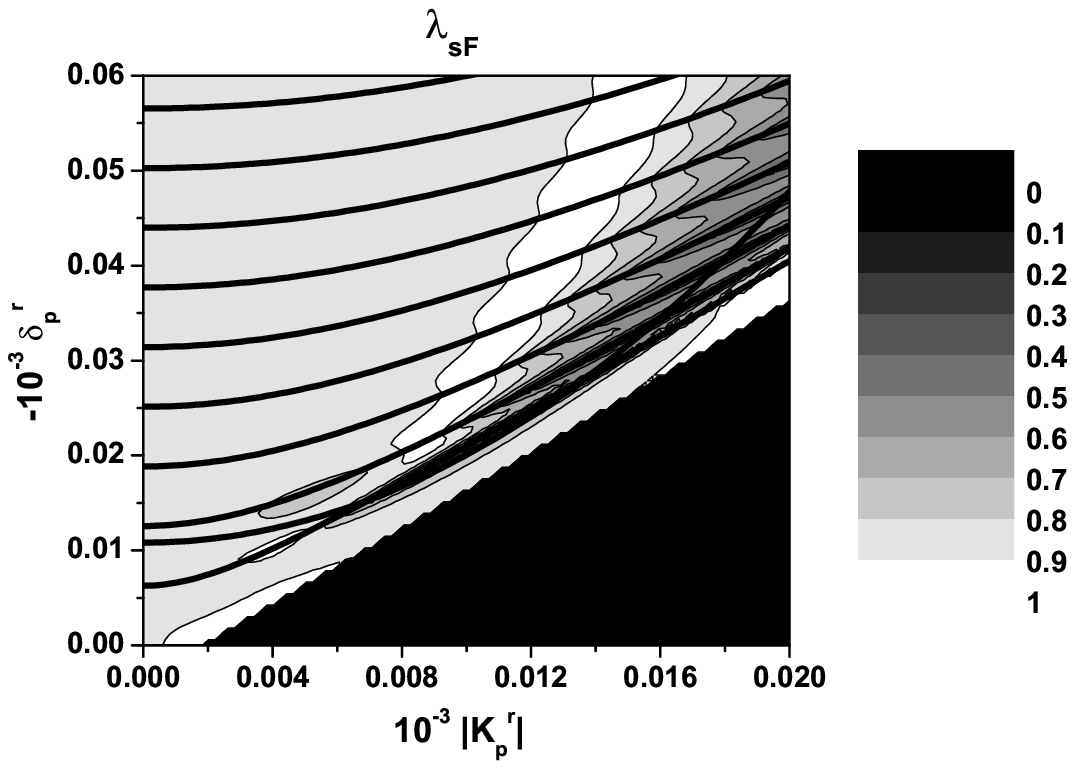}}

 \vspace{0mm}
 \raisebox{0 cm}{b)} \hspace{5mm}
 \resizebox{1.\hsize}{!}{\includegraphics{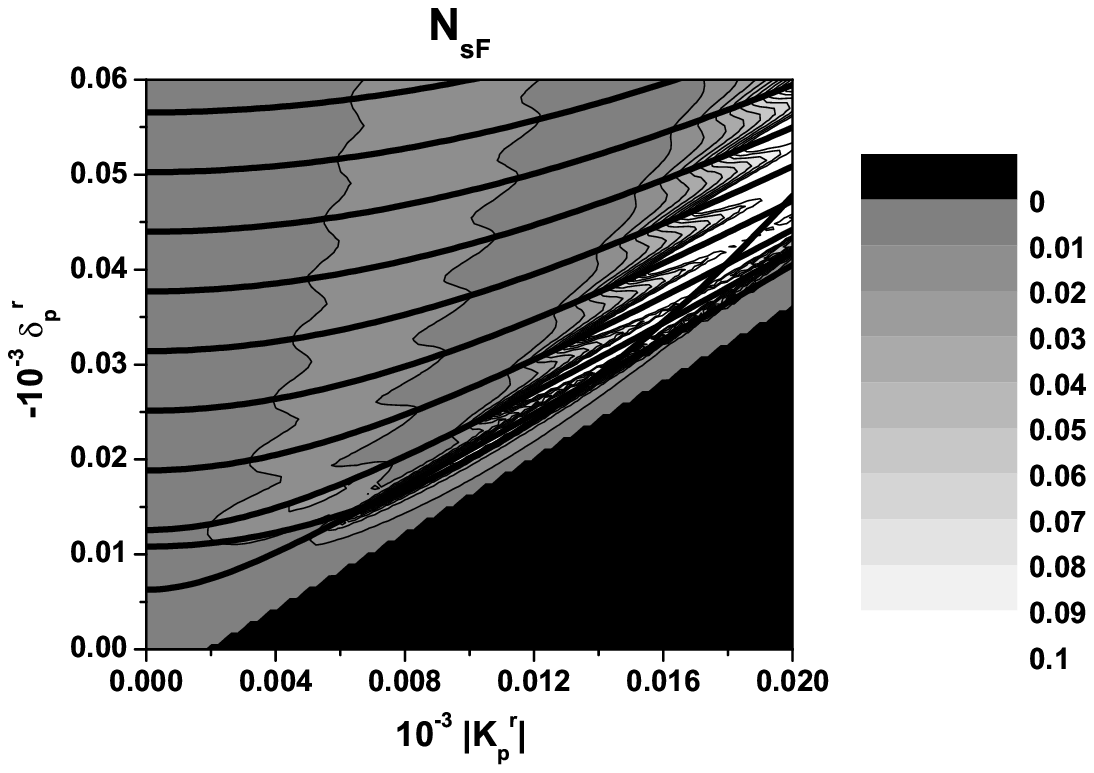}}}
 \vspace{0mm}

 \caption{Contour plots of principal squeeze variance $ \lambda_{s_F} $ (a)
  and number $ N_{s_F} $ of photons leaving the waveguide
  (b) for forward-propagating second-subharmonic field
  as they depend on parameters $ |K_p^r| $ and $ \delta_p^r $ of the
  corrugation in the pump field; more-less equidistant curves in
  the graphs indicate positions of transmission peaks
  [Eq.~(\ref{40})], the last curve going up is given by the condition in
  Eq.~(\ref{35});
  $ \Lambda_{nl}^r = 3.53 \times 10^{-3} $ ($ \delta_{nl,1}^r = - 10.82 $),
  $ \arg(K_p^r) = \pi/2 $;
  values of the other parameters are the same as in Fig.~\ref{fig4}.
  Triangles in lower right corners with formally zero values in both graphs
  lie in the region with an exponential behavior
  of classical amplitudes that is not suitable for nonclassical-light generation.}
\label{fig5}
\end{figure}

A typical dependence of the principal squeeze variance $
\lambda_{s_F} $ as well as the number $ N_{s_F} $ of photons
leaving the waveguide for forward-propagating second-subharmonic
field for attainable values of parameters of the corrugation is
shown in Fig.~\ref{fig5} assuming a fixed value of the nonlinear
phase mismatch $ \delta_{nl,1}^r $ equal to - 10.82 ($
\Lambda_{nl}^r = 3.53 \times 10^{-3} $), i.e. it corresponds
roughly to the second local maximum of $ \lambda_{s_F} $ in the
curve in Fig.~\ref{fig4}. We can clearly see that an efficient
nonlinear interaction occurs in strips that correspond to
transmission peaks; the larger the number $ m $ of a transmission
peak [see Eq.~(\ref{40})] the weaker the effective nonlinear
interaction. The necessity to fulfill also the condition for
perfect phase matching given in Eq.~(\ref{35}) is evident. The
principal squeeze variance $ \lambda_{s_F} $ reaches values around
0.2 inside the strips around the first several transmission peaks.
The larger the number $ m $ of a transmission peak, the greater
values of linear coupling constant $ K_p^r $ and linear phase
mismatch $ \delta_p^r $ have to be used to reach high levels of
squeezing. Up to several forward-propagating photons can be
present inside the waveguide (see Fig.~\ref{fig5}b) at a given
time instant. This means that the power of the outgoing field is
of the order of $ 10^{-8} $~W (energy of one second-subharmonic
photon inside the waveguide of thickness $ t = 5 \times 10^{-7}
$~m and length $ L = 1 \times 10^{-3} $~m corresponds to the
output power of $ 2.58 \times 10^{-8} $~W). Only the first and the
second transmission peaks can give reasonable values of the power
of the outgoing field.

\begin{figure}    
 \resizebox{0.6\hsize}{!}{\includegraphics{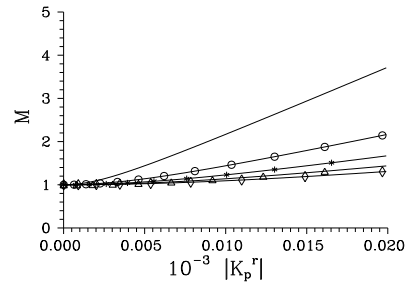}}
 \vspace{2mm}

 \caption{Enhancement factor $ {\cal M} $ as a function of linear
 coupling constant $ K_p^r $ for the first five transmission
 peaks: $ m=1 $ (solid curve), 2 (solid curve with $ \circ $), 3
 (solid curve with $ \ast $), 4 (solid curve with $ \triangle $),
 and 5 (solid curve with $ \diamond $); $ t = 5 \times 10^{-7}
 $~m.}
\label{fig6}
\end{figure}

Effect of a periodic corrugation to the enhancement of the
nonlinear interaction can be approximately quantified as follows.
Because the pump field is strong, its depletion by the nonlinear
process can be omitted and its amplitude $ A_{p_F}(z) $ along the
waveguide is given by the formula in Eq.~(\ref{29}). In our case,
sign + in the condition in Eq.~(\ref{34}) is valid and this means
that the nonlinear process exploits efficiently the first term in
Eq.~(\ref{29}) that is multiplied by the coefficient $ {\cal
B}_{p_F}^+ $. An effective enhancement of the pump-field amplitude
$ A_{p_F} $ inside the waveguide can be given by the ratio $ {\cal
B}_{p_F}^+ / A_{p_F}(0) $ that we call an enhancement factor $
{\cal M} $. The enhancement factor $ {\cal M} $ can be expressed
as follows provided that the pump field lies in a transmission
peak:
\begin{equation}   
 {\cal M} = \frac{2\Delta_p + \delta_p}{4\Delta_p} =
 \frac{1}{2} + \frac{1}{2} \sqrt{ \frac{|K_p|^2+(m\pi/L)^2 }{
 (m\pi/L)2} }.
\label{48}
\end{equation}
According to Eq.~(\ref{48}) the greatest value of enhancement
factor $ {\cal M} $ occurs at the first transmission peak ($ m=1
$) and the greater the linear coupling constant $ K_p $ the
greater the value of enhancement factor $ {\cal M} $. The
enhancement factor $ {\cal M} $ for the first five transmission
peaks is shown in Fig.~\ref{fig6} as a function of the linear
coupling constant $ K_p^r $. The greatest enhancement of an
electric-field amplitude in the first transmission peak is
accompanied by the greatest difference between the maximum and
minimum values of electric-field intensities along the waveguide.
This poses the question whether other types of distributed
feedback resonators (like a quarter wave shifted distributed
feedback resonator) giving a more uniform distribution of
electric-field intensities along the waveguide \cite{Wang1999} can
lead even to a better enhancement of the nonlinear process.
Modelling of such distributed feedback resonators is, however,
beyond the scope of the developed quantum consistent model and so
we keep this question open.

It is useful to compare the achievable values of the enhancement
factor $ {\cal M} $ with those typical for cavity geometries.
Assuming a symmetric planar cavity with mirrors having intensity
reflection $ R $, an electric-field amplitude $ A^{\rm cav} $
inside a cavity is determined from an incident electric-field
amplitude $ A^{\rm ini} $ along the formula $ A^{\rm cav} = A^{\rm
ini} / \sqrt{1-R} $. High quality cavities can have $ R = 99.9
$~\% and so the amplitude $ A^{\rm cav} $ is enhanced with respect
to the amplitude $ A^{\rm ini} $ by factor 30, i.e. the
enhancement of electric-field amplitudes is considerably greater
in this case. On the other hand, the effective nonlinearity
increases also due to field confinement in the transverse plane in
a waveguide. 

Optimum values of waveguide parameters with respect to
squeezed-light generation can be determined along the following
procedure:
\begin{itemize}
 \item The greatest possible value of depth $ t_l $ of a periodic
  corrugation should be used to maximize the pump-field scattering.
  This gives the value of linear coupling constant $ K_p $.
  From practical point of view, the depth $ t_l $ of periodic
  corrugation is limited by technological reasons (also
  validity of the model would have to be judged for deeper
  corrugations).
 \item The value of linear phase mismatch $ \delta_p^r $ should be
  determined along the relation in Eq.~(\ref{41}) to insure that
  the pump field is in the first transmission peak. Both signs of
  the linear phase mismatch $ \delta_p $ are possible and the corresponding
  period $ \Lambda_l $ of linear corrugation is determined according to
  Eq.~(\ref{10}).
 \item Solution of Eq.~(\ref{43}) gives an appropriate value of the
  nonlinear phase mismatch $ \delta_{nl,q} $ under the following
  requirements: sign + in Eq.~(\ref{43}) is used and sign of the
  determined nonlinear phase mismatch $ \delta_{nl,q} $ has to be
  the same as that of the linear phase mismatch $ \delta_p $
  determined in the previous step.
 \item Numerical analysis in the surroundings of the analytically-found values
  of parameters $ \delta_p $ and $ \delta_{nl,\pm 1} $ finally gives
  the values of waveguide parameters optimum for squeezed-light
  generation.
\end{itemize}

\begin{figure}    
 {\raisebox{3.3 cm}{a)} \hspace{5mm}
 \resizebox{0.65\hsize}{!}{\includegraphics{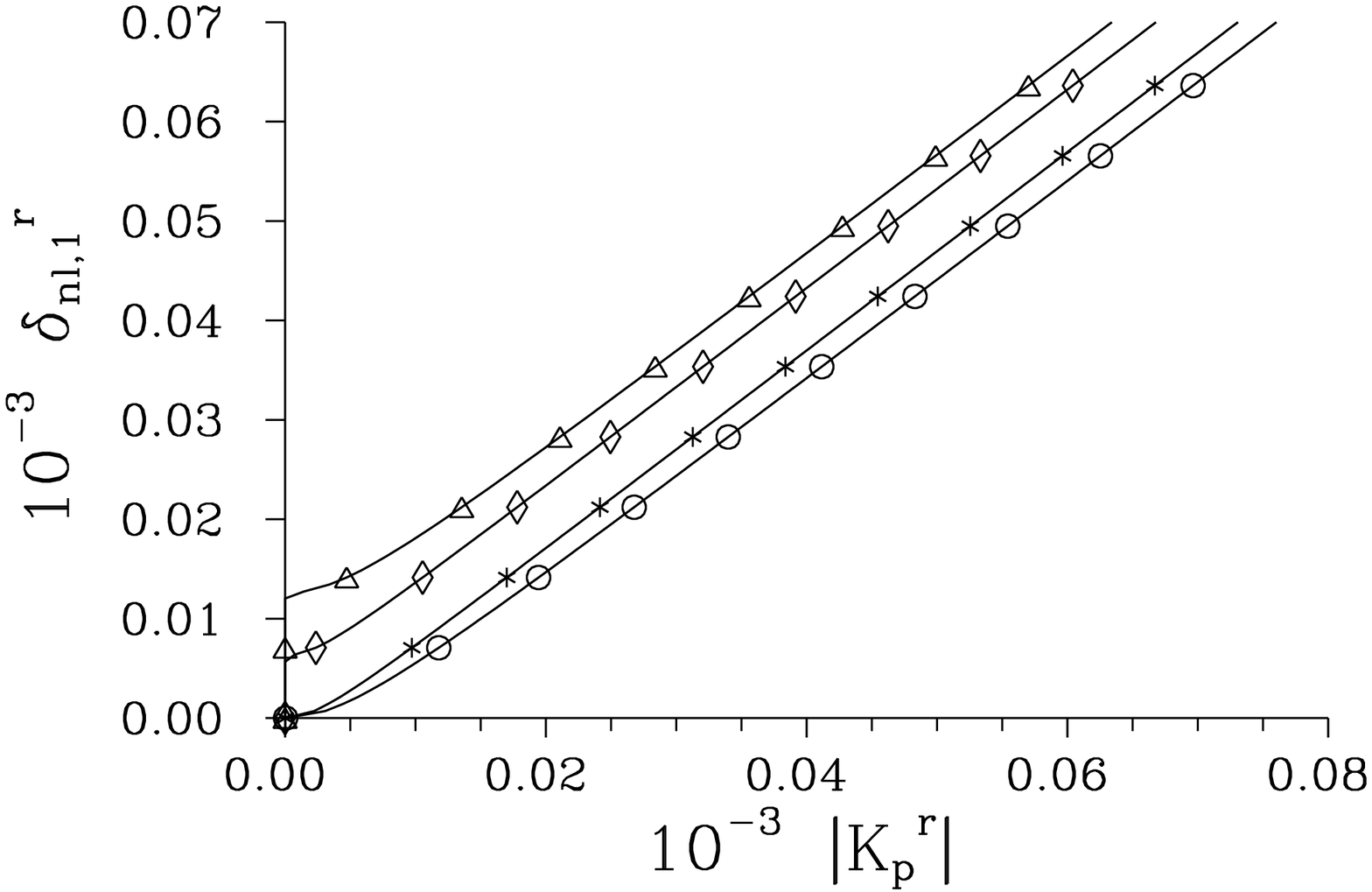}}

 \vspace{0mm}
 \raisebox{3.3 cm}{b)} \hspace{8mm}
 \resizebox{0.6\hsize}{!}{\includegraphics{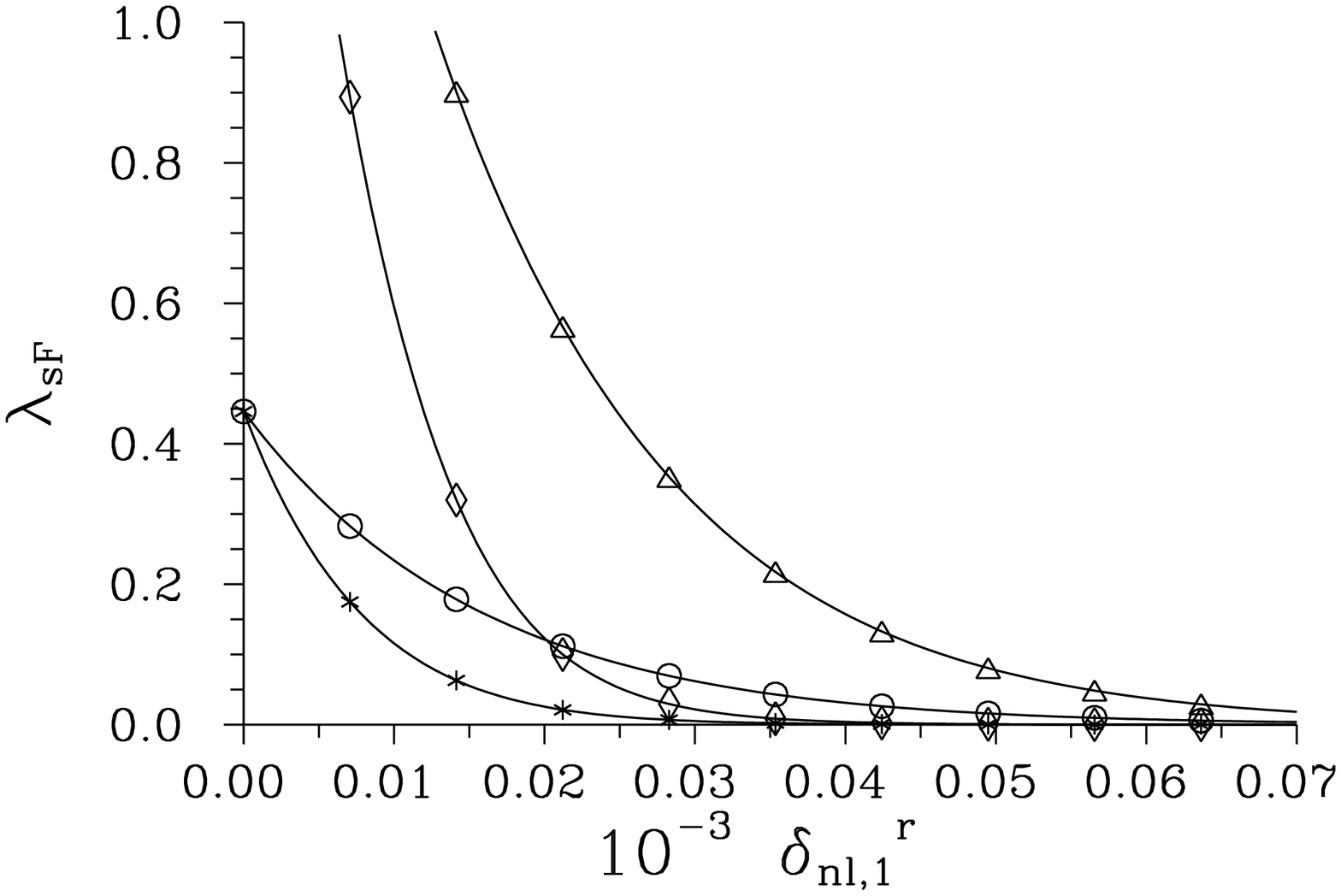}}
}
 \vspace{0mm}

 \caption{Nonlinear phase mismatch $ \delta_{nl,1}^r $ as a
  function of the absolute value $ |K_p^r| $ of linear coupling constant [Eq.~(\ref{43})]
  (a) and principal squeeze variance $ \lambda_{s_F} $ as it depends on the
  nonlinear phase mismatch $ \delta_{nl,1}^r $ (b); first (m=1) and second
  (m=2) transmission peaks as well as  signs +  and - in Eq.~(\ref{43})
  are considered [$ m=1+ $ (solid curve with $ \ast $), $ m=1- $ (solid curve with
  $ \diamond $), $ m=2+ $ (solid curve with $ \circ $), and $ m=2- $ (solid curve with
  $ \triangle $)]; $ \arg(K_p^r) = \pi/2 $;
  values of the other parameters are the same as in Fig.~\ref{fig4}.}
\label{fig7}
\end{figure}

This procedure is documented in Fig.~\ref{fig7}. The dependence of
the optimum value of nonlinear phase mismatch $ \delta_{nl,1}^r $
on the linear coupling constant $ K_p^r $ is shown in
Fig.~\ref{fig7}a. The greater the value of linear coupling
constant $ K_p^r $ the greater the optimum value of nonlinear
phase mismatch $ \delta_{nl,q}^r $. The corresponding values of
the principal squeeze variance $ \lambda_{s_F} $ are depicted in
Fig.~\ref{fig7}b. The larger the value of nonlinear phase mismatch
$ \delta_{nl,1}^r $ the better the values of the principal squeeze
variance $ \lambda_{s_F} $. Curves in Fig.~\ref{fig7}b also
indicate that sign + in Eq.~(\ref{43}) is appropriate and the
first transmission peak gives the best values of principal squeeze
variance $ \lambda_{p_F} $. Keeping the incident pump-field power
fixed, the depth $ t_l $ of periodic corrugation limits the
achievable values of principal squeeze variance $ \lambda_{s_F} $.

\begin{figure}    
 {\raisebox{3.3 cm}{a)} \hspace{5mm}
 \resizebox{0.6\hsize}{!}{\includegraphics{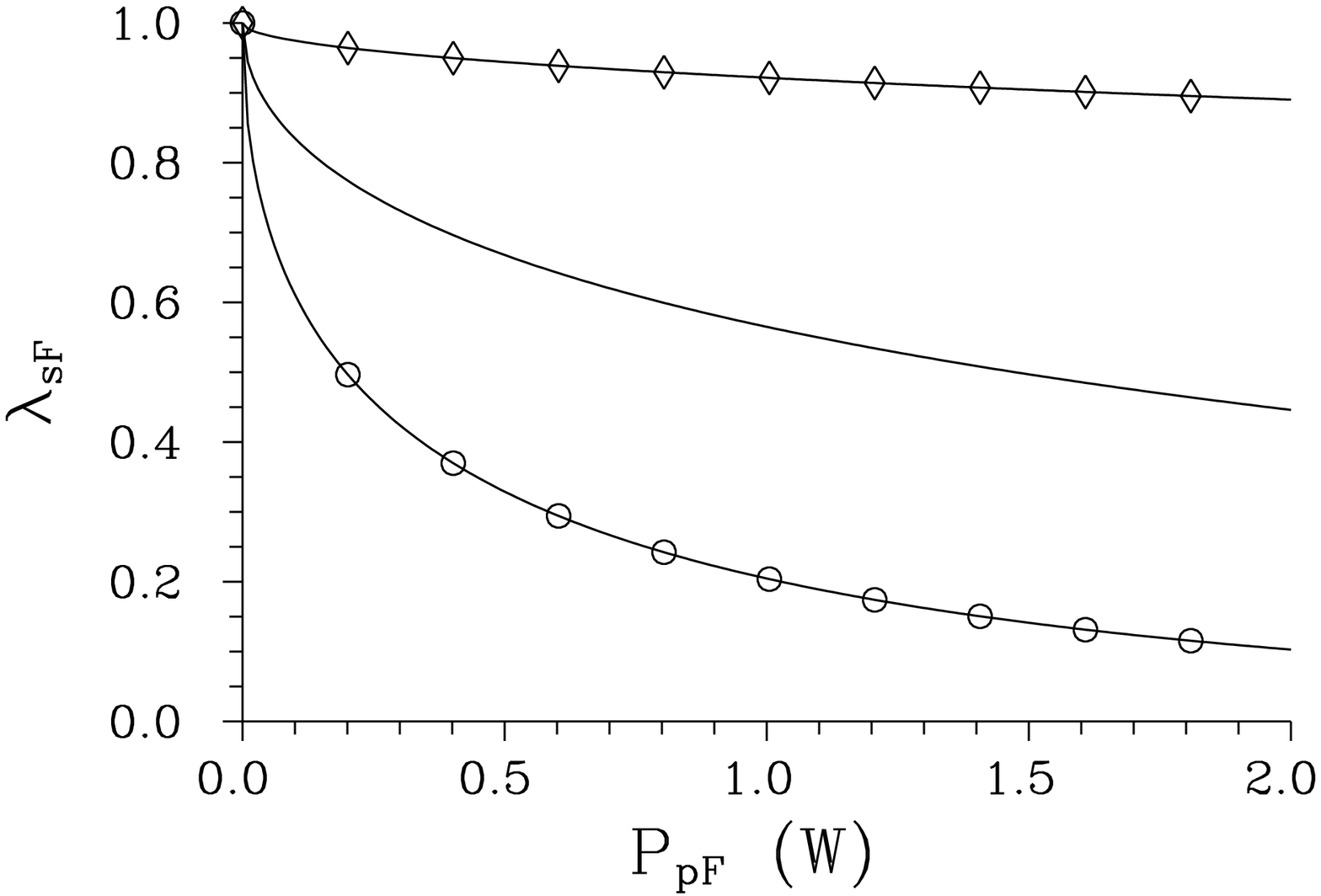}}

 \vspace{0mm}
 \raisebox{3.3 cm}{b)} \hspace{5mm}
 \resizebox{0.6\hsize}{!}{\includegraphics{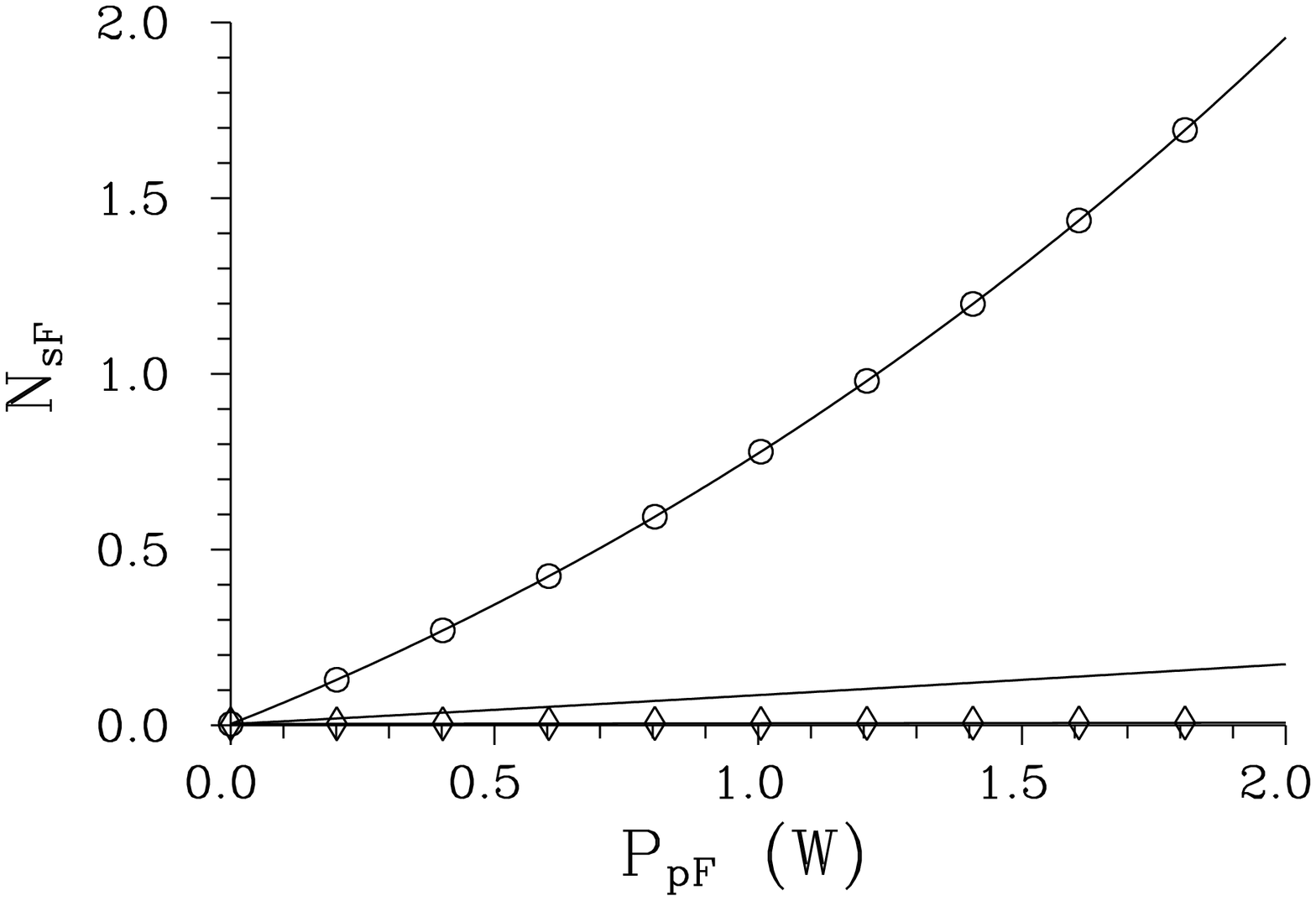}}}
 \vspace{0mm}

 \caption{Principal squeeze variance $ \lambda_{s_F} $ (a) and number $ N_{s_F} $
 of photons leaving the waveguide for forward-propagating second-subharmonic
 field as they depend on the incident pump-field power $ P_{p_F} $
 under different conditions: perfect quasi-phase matching
 ($ \delta_{nl,1}^r = 0 $) without corrugation (solid curve),
 quasi-phase matching with $ \Lambda_{nl}^r = 3.53 \times
 10^{-3} $ ($ \delta_{nl,1}^r = - 10.82 $)
 without corrugation (solid curve with $ \diamond $), and quasi-phase matching
 with $ \Lambda_{nl}^r = 3.53 \times 10^{-3} $ together with a corrugation
 [its parameters are given by the conditions in Eqs.~(\ref{43}) ($ m=1 $)
 and (\ref{35})] (solid curve with $ \circ $);
 $ \arg(K_p^r) = \pi/2 $;
 values of the other parameters are the same as in Fig.~\ref{fig4}.}
\label{fig8}
\end{figure}

To see usefulness of the corrugation we compare two
configurations: a perfectly quasi-phase matched waveguide and a
non-perfectly quasi-phase matched waveguide with a suitable
corrugation that compensates for the given phase mismatch. The
waveguide with corrugation gives better values of the principal
squeeze variance $ \lambda_{s_F} $ and also considerably greater
values of the number $ N_{s_F} $ of photons leaving the waveguide,
as documented in Fig.~\ref{8}. Also improvement caused by an
introduction of corrugation into a non-perfectly
quasi-phase-matched waveguide is worth mentioning (see
Fig.~\ref{fig8}).

\begin{figure}    
 \resizebox{0.6\hsize}{!}{\includegraphics{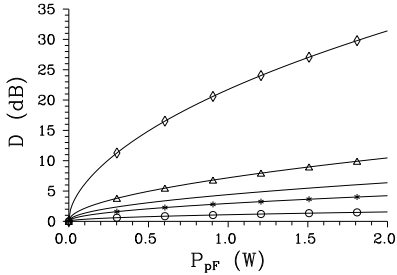}}
 \vspace{2mm}

 \caption{Coefficient $ {\cal D} $ giving the improvement of squeezing
 as it depends on the incident pump power $ P_{p_F} $ for
 $ |K_p^r| = 5 $ (solid curve with $ \circ $), $ |K_p^r| = 10 $
 (solid curve with $ \ast $), $ |K_p^r| = 20 $
 (solid curve with $ \triangle $), and $ |K_p^r| = 50 $
 (solid curve with $ \diamond $). Solid curve without symbols is drawn
 for $ \delta_{nl,1}^r = -10.82 $. Tuning to the first transmission
 peak ($ m=1 $) as well as the condition for overall phase matching
 are assumed; $ t = 5 \times 10^{-7} $~m, $ K_s^r = 0 $,
 $ \arg(K_p^r) = \pi/2 $, $ P_{s_F} = 1 \times 10^{-10} $~W,
 $ \arg(A_{s_F}) = 0 $, $ \arg(A_{p_F}) = 0 $,
 $ A_{s_B} = A_{p_B} = 0 $.}
\label{fig9}
\end{figure}

Benefit of the corrugation to squeezed-light generation can be
quantified defining coefficient $ {\cal D} $ that gives the ratio
(in dB) of the principal squeeze variance $ \lambda_{s_F} $
reached with a corrugation and the principal squeeze variance $
\lambda_{s_F}^{\rm ref} $ characterizing a perfectly
quasi-phase-matched waveguide without corrugation:
\begin{equation}   
 {\cal D} = - 10 \log_{10} \left( \frac{ \lambda_{s_F} }{
  \lambda_{s_F}^{\rm ref} } \right) ;
\label{49}
\end{equation}
$ \log_{10} $ stands for decimal logarithm. The value of
coefficient $ {\cal D} $ determined under the optimum values of
parameters of the corrugation as given in Sec.~III increases with
an increasing incident pump-field power $ P_{p_F} $ (see
Fig.~\ref{fig9}). According to curves in Fig.~\ref{fig9} the use
of periodic corrugation leads to a significant improvement of
squeezing, especially for deeper corrugations and greater incident
pump powers.

The role of length $ L $ of the waveguide can be addressed using
the above-mentioned results as follows. Because the nonlinear
phase mismatch $ \delta_{nl,1} $ does not depend on the length $ L
$, the dimensionless nonlinear phase mismatch $ \delta_{nl,1}^r $
depicted in Fig.~\ref{fig7}a is linearly proportional to the
length $ L $ and, according to curves in Fig.~\ref{fig7}b, the
larger the value of length $ L $ the lower the value of the
principal squeeze variance $ \lambda_{s_F} $. The analysis of
conditions in Eqs.~(\ref{41}) and (\ref{43}) for optimum
squeezed-light generation shows that $ \delta_{nl,1} \rightarrow
|K_p| $ and $ \delta_p \rightarrow 2|K_p| $ in the limit of large
length $ L $. We note that the distance between the curves
corresponding to adjacent transmission peaks in contour plots
depicted in Fig.~\ref{fig5} behaves as $ 1/L $.

A periodic corrugation can be alternatively introduced into the
second-subharmonic field. The results obtained for principal
squeeze variances $ \lambda $ and numbers $ N $ of photons are
comparable to those achieved with a corrugation in the pump field
owing to an increase of electric-field amplitudes of the
second-subharmonic field inside the waveguide. Suitable conditions
are given in Eqs.~(\ref{38}) and (\ref{40}) in this case.

\subsection{Second-harmonic generation}

Second-harmonic generation requires a strong incident
second-subharmonic field. As a reference we consider an incident
power $ P_{s_F} $ of the second-subharmonic field to be 2~W and
periodical poling giving perfect quasi-phase matching. We then
have for the principal squeeze variances $ \lambda_{s_F} \approx
0.87 $ and $ \lambda_{p_F} \approx 0.99 $ ($ \Lambda_{nl}^r
\approx 3.552 \times 10^{-3} $).

The nonlinearly interacting fields behave similarly as in the case
of second-subharmonic generation. If we introduce a periodic
corrugation into the second-subharmonic field and set the
nonlinear phase mismatch $ \delta_{nl,1}^r $ equal to $ 3.53
\times 10^{-3} $, values of the principal squeeze variance $
\lambda_{s_F} $ approach 0.6 inside the strips in the plane
spanned by variables $ |K_s^r| $ and $ \delta_s^r $ where the
conditions from Eqs.~(\ref{38}) and (\ref{40}) are fulfilled
(compare Fig.~\ref{fig5}). The number $ N_{s_F} $ of
second-subharmonic photons at the output of the waveguide
(together with the output power) decreases by an order of
magnitude inside these strips compared to the incident power as a
consequence of transfer of energy from the strong
second-subharmonic field into the pump field (due to an efficient
nonlinear interaction) and also transfer of energy into the
backward-propagating field (due to scattering) is considerable.
Despite this the squeezed second-subharmonic field remains very
strong, it contains about $ 10^{6} $ photons inside the waveguide.
The pump field that is only weakly squeezed for perfect
quasi-phase matching can reach values of the principal squeeze
variance $ \lambda_{p_F} $ around 0.8 assuming a corrugation with
parameters given by Eqs.~(\ref{38}) and (\ref{40}). The pump field
gets a considerable amount of energy from the second-subharmonic
field and so typical values of the number $ N_{p_F} $ of pump
photons leaving the waveguide can reach $ 10^6 $; i.e. the output
power is of the order of $ 10^{-1} $~W (energy of one pump photon
inside the waveguide of thickness $ t = 5 \times 10^{-7} $~m and
length $ L=1 \times 10^{-3} $~m corresponds to the output power of
$ 5.13 \times 10^{-8} $~W).

A periodic corrugation can also be put into the pump field and we
have qualitatively the same results as if the corrugation is
present in the second-subharmonic field. Assuming $
\delta_{nl,1}^r = -10.82 $ as above, values of the principal
squeeze variance $ \lambda_{s_F} $ can reach even 0.3. On the
other hand, values of the principal squeeze variance $
\lambda_{p_F} $ lie above 0.9.

The influence of phases of the interacting optical fields to the
nonlinear process is of interest. It can be easily shown that any
solution of Eqs.~(\ref{8}) depends only on the phase $ \psi =
\arg(K_p) - 2\arg(K_s) $. In our numerical investigations, we did
not observe any dependence of numbers $ N $ of photons as well as
principle squeeze variances $ \lambda $ on the phase $ \psi $. On
the other hand, these quantities depend weakly on phases $
\arg(A_{s_F}) $ and $ \arg(A_{p_F}) $ of the incident fields.
However, this dependence is very weak under the conditions where a
strongly squeezed light is generated.

At the end, we compare values of squeezing achievable in the
considered waveguide with those reached in the commonly used
cavity geometries. The best achieved values of squeezing generated
in a cavity approach to 10~dB \cite{Vahlbruch2007} below the
shot-noise level at present, i.e. values of the principal squeeze
variance $ \lambda_{s_F} $ lie slightly above 0.1. The analyzed
waveguide does not reach so low values of the principal squeeze
variance $ \lambda_{s_F} $ ($ \lambda_{s_F} \approx 0.2 $)
because, as pointed out above, the enhancement of the pump-field
amplitude inside the nonlinear medium is considerably lower
compared to high quality cavities. On the other hand, the
considered waveguide is relatively broad along the $ y $ axis ($
\Delta y = 1\times 10^{-5} $~m) and so narrowing of the waveguide
is possible. This would lead to greater values of the effective
nonlinearity and subsequently to better values of squeezing. Also
longer waveguides can be considered. We believe that the analyzed
waveguide with a periodic corrugation has potential to deliver
squeezed light with values of parameters comparable to those
measured in cavity geometries.

\section{Conclusions}

We have shown that an additional scattering of two nonlinearly
interacting optical fields caused by a small periodic corrugation
on the surface of the waveguide can lead to an enhancement of the
nonlinear process thus resulting in higher generation rates and
better values of squeezing. Origin of this enhancement lies in
constructive interference of the scattered fields leading to
higher values of electric-field amplitudes inside the waveguide.
Optimum conditions for this enhancement have been found
approximately analytically and confirmed numerically. To observe
this effect the natural nonlinear phase mismatch has to be nonzero
in order to match with periodic oscillations caused by scattering
at the corrugation. The deeper the corrugation and the higher the
incident pump power the lower the values of the principal squeeze
variances. A periodic corrugation can be designed to match either
the pump or the second-subharmonic field, or even both of them.

The obtained results have shown that nonlinear planar waveguides
with a periodically corrugated surface represent a promising
source of squeezed light for integrated optoelectronics of near
future.

\appendix

\section{Modes of an anisotropic waveguide}

A mode of the considered waveguide \cite{Yeh1988} depicted in
Fig.~\ref{fig1} is given as a solution of the wave equation
written in Eq.~(\ref{7}). The waveguide is made of LiNbO$ {}_3 $
crystal using the method of proton exchange. The crystallographic
$ z $ axis coincides with the $ x $ axis of the coordinate system
(see Fig.~\ref{fig1}). Ordinary ($ n_{s,o} $) and extraordinary ($
n_{s,e} $) indices of refraction of LiNbO$ {}_3 $ valid for the
substrate and used in calculations are given as:
\begin{eqnarray}     
 n_{s,a}^2 &=& A_a + \frac{B_a}{\lambda^2-C_a} -
 D_a\lambda^2,
 \hspace{1cm} a=o,e, \nonumber \\
 & & A_o = 4.91300, B_o = 0.118717, \nonumber \\
 & & C_o = 0.045932, D_o = 0.0278, \nonumber \\
 & & A_e = 4.57906, B_e = 0.099318, \nonumber \\
 & & C_e = 0.042286, D_e = 0.0224;
\end{eqnarray}
wavelength $ \lambda $ is in $ \mu $m. After proton exchange,
ordinary ($ n_{w,o} $) and extraordinary ($ n_{w,e} $) indices of
refraction of LiNbO$ {}_3 $ characterizing the waveguide are
reached:
\begin{eqnarray}     
 n_{w,o} &=& n_{s,o} - \frac{1}{3} \delta n , \nonumber \\
 n_{w,e} &=& n_{s,e} + \delta n , \nonumber \\
 (\delta n)^2 &=& A_1 + \frac{B_1}{\lambda^2-C_1} -
 D_1 \lambda^2; \nonumber \\
 & & A_1 = 0.007596, B_1 = 0.001129, \nonumber \\
 & & C_1 = 0.116926, D_1 = - 0.0003126.
\end{eqnarray}
We assume that air is present above the waveguide, i.e.:
\begin{equation}    
 n_u = 1.
\end{equation}

Because only the extraordinary index of refraction of LiNbO$ {}_3
$ increases during proton exchange, only TM waves can be guided.
For this reason, instead of solving Eq.~(\ref{7}) for $ x $ and $
z $ components of the electric-field mode functions $ {\bf e}_a $,
we solve the following equation for the only nonzero $ y $
component of the magnetic-field mode functions $ {\bf h}_a(x) $
(fields are assumed to be homogeneous along the $ y $ axis)
\cite{Snyder1983}:
\begin{eqnarray}    
 \frac{d^2{\bf h}_a(x)}{dx^2} + \left[ - \frac{\bar{\bf
 \epsilon}_{zz}(x,\omega_a)}{\bar{\bf \epsilon}_{xx}(x,\omega_a)}
 \beta_a^2 + \frac{\bar{\bf
 \epsilon}_{zz}(x,\omega_a)\omega_a^2}{c^2} \right] {\bf h}_a(x) =
 0. \nonumber \\
 & &
\label{A4}
\end{eqnarray}
The solution of Eq.~(\ref{A4}) for $ y $ component of the
magnetic-field mode function $ {\bf h}_a(x) $ ($ a=p,s $) can be
written as:
\begin{eqnarray}     
 [{\bf h}_a(x)]_{y} &=& - C_a \frac{h_a}{\tilde{q}_a} \exp(-qx), \hspace{2mm}
 x>0; \nonumber \\
 &=& C_a \left[ - \frac{h_a}{\tilde{q}_a} \cos(h_a x) + \sin(h_a x) \right],
  \hspace{2mm} 0<x< -t ; \nonumber \\
 &=& -C_a \left[ \frac{h_a}{\tilde{q}_a} \cos(h_a t) + \sin(h_a t) \right]
 \exp(p_a t)  \nonumber \\
 & & \mbox{} \times \exp(p_a x), \hspace{2mm} x<-t;
\label{A5}
\end{eqnarray}
where $ C_a $ denotes a normalization constant. We have the
following expressions for the coefficients $ h_a $, $ q_a $, $ p_a
$, $ \tilde{p}_a$ and $ \tilde{q}_a $ for the considered
orientation of LiNbO$ {}_3 $:
\begin{eqnarray}   
 h_a &=& \sqrt{ \left(\frac{n_{w,o}(\omega_a)\omega_a}{c}\right)^2 -
  \left( \frac{n_{w,o}(\omega_a)}{n_{w,e}(\omega_a)} \beta_a \right)^2}, \nonumber \\
 q_a &=& \sqrt{ \beta_a^2 - \left(\frac{n_{u}\omega_a}{c}\right)^2}, \nonumber \\
 p_a &=& \sqrt{ \left( \frac{n_{s,o}(\omega_a)}{n_{s,e}(\omega_a)}\beta_a \right)^2 -
  \left(\frac{n_{s,o}(\omega_a)\omega_a}{c}\right)^2},
 \nonumber \\
 \tilde{p}_a &=& \frac{n_{w,o}^2(\omega_a)}{n_{s,o}^2(\omega_a)} p_a, \nonumber \\
 \tilde{q}_a &=& \frac{n_{w,o}^2(\omega_a)}{n_{u}^2} q_a.
\label{A6}
\end{eqnarray}
The solution written in Eq.~(\ref{A5}) holds provided that the
following dispersion relation giving a propagation constant $
\beta_a $ as a function of frequency $ \omega_a $ is fulfilled:
\begin{equation}    
 \tan(h_a t) =
 \frac{h_a(\tilde{p}_a+\tilde{q}_a)}{h_a^2-\tilde{p}_a\tilde{q}_a}.
\label{A7}
\end{equation}
We note that possible solutions of Eq.~(\ref{A7}) for $ \beta_a $
lie in the interval $ n_{s,e}(\omega_a) \omega_a /c \le \beta_a
\le n_{w,e}(\omega_a) \omega_a /c $.

Components of the electric-field mode function $ {\bf e}_a(x) $
can be derived from the magnetic-field mode functions $ {\bf
h}_a(x) $ along the relations:
\begin{eqnarray}    
 [{\bf e}_a(x)]_{x} &=& \frac{\beta_a}{\omega_a \epsilon_0
 \bar{\bf \epsilon}_{xx}(x,\omega_a)} [{\bf
  h}_a(x)]_{y}, \mbox{} \nonumber \\
 \mbox{} [{\bf e}_a (x)]_{y}  &=& 0, \nonumber \\
 \mbox{} [{\bf e}_a(x)]_{z} &=& - \frac{i}{\omega_a
 \epsilon_0\bar{\bf \epsilon}_{zz}(x,\omega_a)} \frac{d [{\bf
  h}_a(x)]_{y} }{d x}.
\label{A8}
\end{eqnarray}

The normalization constants $ C_a $ occurring in Eqs.~(\ref{A5})
are determined from the condition that the mode functions $ {\bf
e}_a(x) $ describe one photon with energy $ \hbar\omega_a $ inside
the waveguide (of length $ L $ and thickness $ \Delta y $):
\begin{eqnarray}   
 & & 2 \epsilon_0 \Delta y L \int_{-\infty}^{\infty} dx
 \left[ \bar{\bf \epsilon}_{xx}(x,\omega_a) \left|
 [{\bf e}_a(x)]_x \right|^2 \right. \nonumber \\
 & & \left. \mbox{} + \bar{\bf \epsilon}_{zz}(x,\omega_a) \left|
 [{\bf e}_a(x)]_z \right|^2 \right] = \hbar\omega_a .
\label{A9}
\end{eqnarray}

The corrugation on the surface causes periodic changes of values
of permittivity $ {\bf \epsilon} $ for $ x \in (0,-t_l) $ (see
Fig.~\ref{fig1}) and we have $ \varepsilon_{\pm 1} = i(\bar{\bf
\epsilon}-1) / (\pi \bar{\bf \epsilon}) $ in this case using
Eqs.~(\ref{2}) and (\ref{3}). If the waveguide is periodically
poled $ {\bf d}_{\pm 1} = - 2i/ \pi {\bf d} $ in Eq.~(\ref{5}) and
the remaining coefficients may be omitted. Using the
electric-field mode functions $ {\bf e}_p $ and $ {\bf e}_s $
determined in Eqs. (\ref{A8}), linear ($ K_s $, $ K_p $) and
nonlinear ($ K_{nl,0} $, $ K_{nl,1} $) coupling coefficients
defined in Eqs. (\ref{11}) and (\ref{12}) can be rearranged into
the form:
\begin{eqnarray}   
 K_a &=& \frac{i\omega_a^2}{2\pi c^2 \beta_a} \left[ \frac{n_{w,e}^2(\omega_a)-1}{n_{w,e}^2(\omega_a)}
  \int_{-t_l}^{0} dx\, \left|[{\bf e}_{a}(x)]_x\right|^2 \right. \nonumber \\
  & & \mbox{} \left. + \frac{n_{w,o}^2(\omega_a)-1}{n_{w,o}^2(\omega_a)}
  \int_{-t_l}^{0} dx\, \left|[{\bf e}_{a}(x)]_z\right|^2 \right] \nonumber \\
 & & \mbox{} \times \left[ \int_{-\infty}^{\infty}dx\,  ( \left|[{\bf e}_{a}(x)]_x\right|^2 +
  \left|[{\bf e}_{a}(x)]_z\right|^2 ) \right]^{-1}, \nonumber \\
  & & \mbox{}  \hspace{5mm} a=p,s ,
\label{A10} \\
 K_{nl,0} &=& \frac{i\omega_s^2}{2c^2\beta_s} \int_{-\infty}^{0}
  dx\, {\bf d} \cdot {\bf e}_p(x) {\bf e}_s^*(x) {\bf e}_s^*(x) \nonumber \\
 & & \mbox{} \times \left[ \int_{-\infty}^{\infty} dx\, ( \left|[{\bf e}_{s}(x)]_x\right|^2 +
  \left|[{\bf e}_{s}(x)]_z\right|^2 ) \right]^{-1}, \nonumber \\
  & &
\label{A11}   \\
 K_{nl,1} &=& - \frac{2i}{\pi} K_{nl,0}.
\label{A12}
\end{eqnarray}

Nonzero coefficients of the nonlinear tensor $ {\bf d} $ of LiNbO$
{}_3 $ used in calculations are the following:
\begin{eqnarray}      
 d_{zzz} &=& - d_{zyy} = - d_{yyz} = 3.1 \times 10^{-12} {\rm mV}^{-1} , \nonumber
 \\
 d_{xyy} &=& d_{xzz} = d_{zzx} = d_{yyx} = 5.87 \times 10^{-12} {\rm mV}^{-1}, \nonumber \\
 d_{xxx} &=& 41.05 \times 10^{-12} {\rm mV}^{-1}.
\end{eqnarray}

\acknowledgments This material is based upon the work supported by
the European Research Office of the US Army under the Contract No.
N62558-05-P-0421. Also support coming from cooperation agreement
between Palack\'{y} University and University La Sapienza in Rome
and project 202/050498 of the Czech Science Foundation are
acknowledged.

\end{document}